\documentclass[10pt]{article}
\usepackage[tbtags]{amsmath}
\usepackage[english]{babel}
\usepackage{hyperref}
\usepackage{amssymb,amsthm}
\usepackage{amsfonts}
\usepackage{graphicx}
\graphicspath{{figs/}}
\usepackage[left=2.5cm,right=2.2cm,top=0.5cm,bottom=1.3cm,includeheadfoot]{geometry}
\usepackage{indentfirst}
\usepackage{color}
\usepackage{cite}
\usepackage{mathtools}
\usepackage{graphicx}
\usepackage{subcaption}
\usepackage{tikz-cd}
\usepackage{enumerate}

\definecolor{cites}{rgb}{0.65, 0.20, 0.46}
\definecolor{chapters}{rgb}{0.25, 0.30, 0.56}
\definecolor{header}{rgb}{0.0, 0.5, 0.64}

\hypersetup{
	colorlinks=true, 
	linkcolor=header, 
	citecolor=cites,
	urlcolor=chapters
}

\usepackage[normalem]{ulem}

\numberwithin{equation}{section}

\renewcommand*{\thefootnote}{\fnsymbol{footnote}}
\begin{document}
	\begin{center}
		{\Large\bf
			Analytic solutions for the Bianchi I universe coupled to several barotropic perfect fluids
		}
		\vskip 5mm
		{\large
			David Brizuela\footnote{contact author: {\tt david.brizuela@ehu.eus}}
			and Sara F. Uria\footnote{contact author: {\tt sara.fernandezu@ehu.eus}}
		}
		\vskip 3mm
		{\sl Department of Physics and EHU Quantum Center, University of the Basque Country UPV/EHU,\\
			Barrio Sarriena s/n, 48940 Leioa, Spain}\\
	\end{center}
	
	\setcounter{footnote}{0}
	\renewcommand*{\thefootnote}{\arabic{footnote}}
	
	\begin{abstract}
		We consider
the Bianchi I geometry coupled to several species of comoving barotropic perfect fluids with a linear equation of state in the context of general relativity.
The solution of the dynamics can be reduced to a quadrature, which can be explicitly performed in certain cases. In particular, we obtain the explicit solution for one species, as well as for two species, given their barotropic indices obey a certain relation. These solutions include and generalize different models studied in the literature. For completeness, we analyze all the different possible signs of the matter energy densities, and we obtain a particularly interesting model in which an exotic species produces a bounce of the scale factor, providing a singularity-free cosmology, and then decays to leave a nonexotic component as the dominant fluid for large volumes.
	\end{abstract}

	\section{Introduction}
	\label{sec:intro}

The analysis of the Bianchi I geometry holds special relevance for a number of reasons. On the one hand, it is the simplest among all the Bianchi models.
In fact, during specific periods of evolution,
when the corresponding potential generated by the spatial curvature
is negligible and thus the dynamics is
dominated by the kinetic terms, the dynamics of the different Bianchi models
can be well described by the evolution of Bianchi I.
On the other hand, the Bianchi I geometry has a cosmological relevance
on its own, as it can be considered as an anisotropic generalization of the flat Friedmann-Lema\^itre-Robertson-Walker (FLRW) metric, which describes our Universe up to a high degree of accuracy.

Nowadays, the numerical resolution of the Einstein equations is a common
practice, and can be used to obtain solutions even for highly nonlinear and complicate physical scenarios. However, such solutions are not exact,
and the derivation of analytic solutions is still of relevance since it can
provide specific insight into details of the geometry.
Concerning the Bianchi I geometry in the context of general relativity,
there are several well-known analytic solutions with different matter content.
The solution corresponding to the vacuum case, known as the Kasner solution \cite{Kasner:1921zz}, is probably the most notable one.
Other significant solutions include the Heckmann-Schücking solution \cite{Heckmann-Schucking}, which corresponds to a matter content of a dust field (pressureless fluid). In fact, this solution has been generalized in Refs.~\cite{Khalatnikov:2003ph, Kamenshchik:2009dt} to include dust, stiff matter, and both positive and negative cosmological constants. Furthermore, there is also a large body of results in the literature regarding Bianchi I models with additional symmetry restrictions. In particular, as previously mentioned, the isotropic case corresponds to the FLRW solution. Moreover, there are also several solutions with a local rotational (axial) symmetry,  considering different types of matter (see, e.g., Refs.~\cite{Thorne:1967,Akarsu:2009bhh, Yadav:2011bj, Singh:2018iii, Singh:2019crl, Leon:2021rcx}). However, without restricting to such symmetric cases, obtaining explicit analytic solutions is challenging,
even though Bianchi I is the simplest spatially anisotropic model.

Taking this into account, in the present paper we will obtain several
explicit solutions for Bianchi I coupled to one or several perfect fluid species,
which will contain some of the aforementioned solutions as particular cases.
More precisely, we will consider that the matter content is given
	by $n$ orthogonal (nontilted) and barotropic perfect fluids
	with a linear equation of state. Exact
	analytic solutions will be derived, and their asymptotic properties
	 --- both towards large volumes as well as towards the singularity --- will be analyzed. On the one hand, we show that
for large volumes all solutions isotropize and they tend to a flat FLRW.
On the other hand, around the singularity, where curvature invariants blow up, they converge to the Kasner dynamics.

	The paper is organized as follows. In Sec.~\ref{sec.bianchimodels} we present the basic variables and the Einstein equations for the Bianchi I
	geometry, with the matter content given by $n$ species
	of orthogonal (nontilted) barotropic perfect fluids. We also introduce the curvature invariants that we will later analyze for the exact solutions. Then, in Sec.~\ref{sec.generaldescription}, we fix the time gauge that will be used all along the paper, and give a general qualitative description of the dynamics based on an energy equation. Once this is set, in Sec.~\ref{sec.solutions}	
	we present analytic solutions for certain specific fluid components. More precisely, we obtain the
	solution for vacuum (Kasner), the general solution for one fluid species, as well as an analytic
	solution for two fluid species provided that their corresponding barotropic indices satisfy a certain relation.
	In all of the cases, we analyze the scenario with different signs
	of the energy densities, which, in particular, leads to
	a singularity-free cosmology for the model with two fluid species. Subsequently, in Sec.~\ref{sec.asymptoticsBI}
	the asymptotic behavior of the solutions is analyzed. And, in Sec.~\ref{sec:conclusions}
	we summarize and discuss the main results of the paper. Finally,
	in the Appendix some plots for the model with two fluids are presented to illustrate the behavior of the different cases.

	\section{Bianchi I geometry coupled to barotropic fluids}\label{sec.bianchimodels}

	\subsection{Basic variables and equations of motion}

The	Bianchi I geometry describes a spatially homogeneous, though anisotropic, universe.
	Its corresponding metric can be generically written as follows:
		\begin{align}\label{metric}
	ds^2=-N^2 dT^2+a_1^2 dx_1^2+a_2^2 dx_2^2+a_3^2 dx_3^2,
	\end{align}
where $a_i=a_i(T)$, for $i=1,2,3$, are the scale factors in the three different spatial directions,
while $N=N(T)$ is the lapse function. Following Misner \cite{Misner:1969ae,Misner:1969hg}, it is convenient to define the shape parameters,
	\begin{align}\label{def_misner_var}
	\beta_{+}&:=-\frac{1}{2}\ln \left[{\frac{a_3}{(a_1a_2a_3)^{1/3}}}\right],\\
	\beta_{-}&:=\frac{1}{2\sqrt{3}}\ln \left({\frac{a_1}{a_2}}\right),
\end{align}
which encode the spatial anisotropy of the universe. In particular, if all of the scale factors are equal,
$a_1=a_2=a_3$, then $\beta_{\pm}=0$. In addition, we also introduce the average scale factor,
\begin{equation}
	a :=(a_1 a_2 a_3)^{1/3},
\end{equation}
and its corresponding Hubble factor,
	\begin{equation}\label{hubble}
	H:=\frac{a^\prime}{a N},
	\end{equation}
	where the prime stands for a derivative with respect to the generic time $T$.
Therefore, the set $(a,\beta_+,\beta_-)$ will be our basic variables to describe the geometric
degrees of freedom.

    Concerning the matter content,
	we will assume that it is given by a collection of $n$ species of comoving perfect fluids
	with energy-momentum tensor
	\begin{equation}
	\label{energy_momentum_tensor}
	T_{\mu\nu}=(\rho+p)u_\mu u_\nu+p g_{\mu\nu},
	\end{equation}
	where $\rho=\sum_{s=1}^n \rho_s$ is the total mass-energy density, $p=\sum_{s=1}^n p_s$ is the total pressure,
	and $u_\mu$ is the velocity of the fluids, which obeys the normalization condition $u^\mu u_\mu=-1$.
	Note that, in general, the fact that all species have the same velocity is an additional assumption
	of the model. For one species $(n=1)$, the present diagonal model \eqref{metric} does not allow for a single tilted fluid,
	since the components of the Einstein tensor $G^0{}_i$ are vanishing, which states the absence of matter current,
	and thus $u_i=0$. However, one could have several tilted fluids with different
	velocities, and, in particular, with nonvanishing spatial velocities, as long as the net
	current vanishes (see, e.g.,
	\cite{Coley:1986, Cembranos:2019plq, Sandin:2008bq} for some examples in FLRW and in Bianchi I).

	The continuity equation for the matter fields can be derived from the conservation of the energy-momentum tensor, $\nabla_\mu T^{\mu}{}_{\nu}=0$, which,
	assuming that the interaction between different species is only gravitational, implies
	the following $n$ relations
	\begin{equation}
	\label{eq_rho_s}
	{\rho_s}^\prime+3(\rho_s+p_s)\frac{a^\prime}{a}=0,\qquad{\rm for}\,\,\, s=1,\dots,n,
	\end{equation}
	where the prime $(^\prime)$ denotes derivation with respect to the generic time $T$.
Finally,
the Einstein equations for this model can then be reduced to the following set of equations:
	\begin{align}
	\label{eq_motion_alpha}
	\frac{a^{\prime\prime}}{a}&=-2\frac{{a^\prime}^2}{a^2}+\frac{\kappa}{2}N^2 (\rho-p)
+\frac{{N}^\prime{a}^\prime}{Na},
	\\
	\label{eq_motion_beta_+}
	{\beta}_+^{\prime\prime}&= -3\frac{a^{\prime}}{a}\beta_+^{\prime} +\frac{N^{\prime}{\beta}_+^\prime}{N},\\
	\label{eq_motion_beta_-}
	\beta_-^{\prime\prime}&= -3\frac{a^{\prime}}{a}\beta^{\prime}_- +\frac{N^{\prime}\beta_-^{\prime}}{N},\\
	\label{eq_constraint}
	0 &= \kappa\rho-\frac{3}{N^2}\left(\frac{{a^\prime}^2}{a^2}-{\beta^\prime_+}^{\!\!2}-{\beta^{\prime}_-}^{\!\!2}\right).
	\end{align}
In particular, if one imposes the isotropic condition $\beta_\pm=0$,
these equations describe the flat FLRW model. In such a case,
the equations for the shape parameters \eqref{eq_motion_beta_+}--\eqref{eq_motion_beta_-} are
trivially fulfilled, \eqref{eq_constraint} corresponds to the Friedmann equation, and
\eqref{eq_motion_alpha} is the acceleration equation.

	\subsection{Equation of state}
	\label{sec:equation_of_state}

	In summary, there are three geometric $(a, \beta_+, \beta_-)$ and $2 n$
	matter $(\rho_s,p_s)$ dynamical variables, with $s=1,\dots,n$, but,
	among the $(4+n)$ equations of motion \eqref{eq_rho_s}--\eqref{eq_constraint} only $(3+n)$ are linearly independent.
	Therefore, as it is well known, in order to close the system, it is necessary to consider an equation
	of state for each fluid, which provides $n$ additional relations. In the following, we will assume
	barotropic fluids with $p_s=\omega_s \rho_s$. Different values of the constant $\omega_s$, known as the barotropic index,
	describe different matter contents of interest. For instance, $\omega_s=0$ corresponds to pressureless dust,
	$\omega_s=1/3$ to relativistic particles, while $\omega_s=1$ defines stiff matter.
	This latter case is equivalent to a massless scalar field, for which
	the speed of sound equals the speed of light, and it corresponds to the maximum allowed value
	for $\omega_s$ that respects causality.
	
	If one requires the dominant and strong energy conditions to be obeyed, $\rho_s$ must be nonnegative and
	$\omega_s$ is restricted to the interval $-\frac{1}{3}\leq \omega_s\leq 1$. However, there are some cases
	of interest that one can consider outside the commented ranges.
	In particular, a cosmological constant $\Lambda$ can be described as a fluid with
	$\omega_s=-1$, and with a contribution $\rho_s=\Lambda/\kappa$ to the energy density, which, if $\Lambda$ is negative, implies a negative $\rho_s$. Other more speculative fluids are also studied in the literature, like cosmic strings $(\omega_s=-1/3)$, domain walls $(\omega_s=-2/3)$, or phantom energy $(\omega_s<-1)$, in some instances considering also their effects
	with negative energy density (see, e.g., Ref.~\cite{Nemiroff:2014gea} for a study in FLRW).
	Moreover, quantum-gravity effects are supposed to resolve the cosmological singularity, and, if one could describe
	those effects as an effective barotropic fluid, it would violate the energy
	conditions. 
	
	Therefore,
	in order to reproduce and generalize some results of the literature,
	we will not strictly
	impose the energy conditions, though we will exclude phantom energy with $\omega_s<-1$
	and fluids that would violate causality with $\omega_s>1$ .
	Thus, the sign of $\rho_s$ will be taken as arbitrary,
	and the $n$ species will be assumed to have different barotropic indices with
	$-1\leq\omega_-<\dots<\omega_s<\dots<\omega_+\leq 1$ bounded by certain $\omega_-$ and $\omega_+$.
	These inequalities will be saturated in the case there is a cosmological constant in the model, and then $\omega_-=-1$, or if there is a stiff-matter species, and then $\omega_+=1$.
	For the 
	case of a single component trivially $\omega_-=\omega_+$.
	
	Now, for such linear equation of state, it is straightforward to solve the continuity
	equation \eqref{eq_rho_s},
	\begin{equation}
	\label{rho_evolution}
	\rho_s=\rho_{0s} a^{-3(1+\omega_s)},
	\end{equation}
	with constant $\rho_{0s}$. This expression provides us with the asymptotic
	tendency of the matter fields and clearly shows that one of the species will be dominant.
	More precisely, since $1+\omega_s\geq 0$, for small volumes close to the singularity ($a\to 0$), then
	$\rho\approx \rho_{0+}a^{-3(1+\omega_+)}$, while for large volumes ($a\to +\infty$) one obtains
	$\rho\approx \rho_{0-}a^{-3(1+\omega_-)}$. For instance, under the presence
	of stiff matter and cosmological constant, the stiff-matter component will dominate near the singularity,
	while the cosmological constant will dominate for large volumes. During the intermediate stages between those two limits, there will be different epochs dominated by different matter types.

	\subsection{Curvature invariants}

	We will be interested in analyzing the asymptotic behavior of curvature. In particular,
	the Kretschmann scalar,
	\begin{equation}
	\label{def_Kretschmann}
	K=R_{\mu\nu\lambda\rho}R^{\mu\nu\lambda\rho},
	\end{equation}
	is usually used to characterize singularities since, in vacuum spacetimes, other geometric scalars, such as
	the Ricci scalar $R$, or the Ricci square $R_{\mu\nu}R^{\mu\nu}$, are exactly vanishing due to the
	field equations.
	Another interesting scalar is the Weyl square,
	\begin{equation}\label{Weylsquare}
	W^2=C_{\mu\nu\lambda\rho}C^{\mu\nu\lambda\rho},
	\end{equation}
	which is argued to encode the spacetime entropy; following the Weyl curvature hypothesis \cite{Penrose1979}, $W^2$
	should be small at the initial singularity, and then grow as the universe expands.
	
	Making use of the Weyl decomposition, the two scalars above can be related as,
	\begin{equation}
	W^2=K-2 R_{\mu\nu}R^{\mu\nu}+\frac{1}{3}R^2.
	\end{equation}
	Furthermore, from the Einstein equations, considering the energy-momentum tensor \eqref{energy_momentum_tensor}, it is immediate to obtain
	\begin{align}
	R_{\mu\nu}R^{\mu\nu}&=\kappa^2\left(3p^2+\rho^2\right),\\
	R&=\kappa\left(\rho-3p\right).
	\end{align}
	Therefore, the Weyl square is given as
	\begin{equation}
	\label{W2K}
	W^2=K-\kappa^2
	\left(3p^2+
	\frac{5}{3}\rho^2+2p\rho\right)=K-\kappa^2
	\sum_{s=0}^n\sum_{l=0}^n
	\left[
	\rho_{0s}\rho_{0l}\left(
	3\omega_s\omega_l+\frac{5}{3}+2\omega_s
	\right)
	 a^{-3(2+\omega_s+\omega_l)}
	\right],
	\end{equation}
	which, towards the singularity scales as
	\begin{equation}
	\label{Weyl_singularity}
	W^2\approx K-\kappa^2\rho_{0+}^2\left(
	3\omega_+^2+\frac{5}{3}+2\omega_+
	\right)
	a^{-6(1+\omega_+)},
	\end{equation}
	while, towards large volumes,
		\begin{equation}
	\label{Weyl_large_vol}
	W^2\approx K-\kappa^2\rho_{0-}^2\left(
	3\omega_-^2+\frac{5}{3}+2\omega_-
	\right)
	a^{-6(1+\omega_-)},
	\end{equation}
	where $\omega_+$ and $\omega_-$ are the value of the maximum and minimum barotropic indices, respectively, and $\rho_{0+}$ and $\rho_{0-}$ their corresponding densities. It is interesting to note that in vacuum, since $R_{\mu\nu}=0$, the Weyl square and the Kretschmann scalar are identical. Making use of the analytic solutions that will be obtained below,
	we will compute the specific scaling of these curvature invariants, and reproduce the general behavior provided in Ref.~\cite{Barrow:2002is}.

	\section{Gauge fixing and general description of the dynamics}\label{sec.generaldescription}
	
		For the subsequent analysis, we will fix the lapse as $N=a^{3}$, and name the time in this
	gauge as $\tau=T$. In some cases we will translate our results to the cosmological time $t$, for which the lapse
	is $N=1$. The relation between these two times reads,
	\begin{align}\label{eq_cosmological_time}
	\left(\frac{dt}{d\tau}\right)^2=a^{6}\,\Rightarrow t=\int d\tau\, a^{3}(\tau),
	\end{align}
	where the global sign has been chosen so that the flow of both times run in the same direction.

	Denoting with a dot the derivative with respect to the time $\tau$,
	in the $\tau$ gauge
	the equations of motion \eqref{eq_motion_alpha}--\eqref{eq_constraint} read
	\begin{align}
	\label{eq_motion_alpha_lapse_BI}
	\frac{\ddot{a}}{a}&=\frac{\dot{a}^2}{a^2}+\frac{\kappa}{2}a^{6}\left(\rho -p\right),
	\\
	\label{eq_motion_beta_+_lapse_BI}
	\ddot{\beta}_+&= 0,\\
	\label{eq_motion_beta_-_lapse_BI}
	\ddot{\beta}_-&= 0,
	\\
	\label{constraint_BI}
	0&=\kappa\rho-\frac{3}{a^6}\left(\frac{\dot{a}^2}{a^2}-\dot{\beta}_+^2-\dot{\beta}_-^2\right),
	\end{align}
 where one has to take into account that $p=\sum_s \omega_s \rho_s$
	and $\rho=\sum_s\rho_s$, with $\rho_s$ given in \eqref{rho_evolution}.
	Equations \eqref{eq_motion_beta_+_lapse_BI} and \eqref{eq_motion_beta_-_lapse_BI} do not depend on the matter content, and lead to a
	linear evolution of the shape parameters in the time $\tau$,
	\begin{align}\label{sol_beta_+_Kasner}
	\beta_+&=k_++p_+ \tau,\\
	\label{sol_beta_-_Kasner}
	\beta_-&=k_-+p_- \tau,
	\end{align}
	with constants $k_+,k_-,p_+,$ and $p_-$.
	The constants $k_\pm$ are pure gauge and can be absorbed in a redefinition
	of the coordinates,\footnote{Specifically, the change of coordinates to perform is $x_1\to x_1 e^{-k_+-\sqrt{3}k_-}$, $x_2\to x_2 e^{-k_++\sqrt{3}k_-}$, and $x_3 \to x_3 e^{2k_+}$.} while $p_\pm$ are the canonical momenta of $\beta_\pm$,
	and completely encode the anisotropy of the model.
	In particular, the isotropic case corresponds to $p_+=0=p_-$.
	
	Using the above result, the constraint \eqref{constraint_BI} reduces now to
	\begin{equation}
	\label{eq_friedmann}
	\frac{\dot{a}^2}{a^2}=(p_+^2+p_-^2)+\frac{\kappa}{3}\rho a^{6}.
	\end{equation}
	This happens to be the Friedmann equation that describes the evolution of the scale factor
	$a$ in a FLRW universe with the matter content given by the density $\rho$ plus
	a massless scalar field with momentum $P:=(p_+^2+p_-^2)^{1/2}$. Therefore, the evolution
	of the average scale factor $a$ in a Bianchi I universe with a given matter content,
	is completely equivalent to the evolution of the scale factor in a FLRW universe
	with the same matter content plus a scalar field with momentum $P$.
	In addition, using \eqref{rho_evolution}, the matter term takes the form 
	\begin{equation}
	\frac{\kappa}{3}\rho a^{6}=\frac{\kappa}{3} \sum_{s}\rho_{0s} a^{3(1-\omega_s)}.
	\end{equation}
	Hence, for the evolution of the average scale factor $a$ given by \eqref{eq_friedmann},
	both the contribution from the anisotropies $P^2$ and from the stiff-matter ($\omega_s=1$) component $\frac{\kappa}{3}\rho_{0{\rm stiff}}$
	scale as $a^0$, and are thus completely indistinguishable.
	Therefore, for clarity of the presentation, we will assume there is no stiff matter in our model and
	all $\omega_s$ are in the range $\omega_s\in [-1,1)$. However, in order to include a stiff-matter component in the solutions that will be presented
	below, one simply needs to perform the replacement $P^2\rightarrow P^2+\frac{\kappa}{3}\rho_{0{\rm stiff}}$.
	
	In order to check qualitatively how the anisotropies change the evolution
	of the model as compared to its isotropic $(P=0)$ counterpart, it is useful
	to define
	\begin{align}\label{def_V_a}
	V(a):=-\frac{a^2}{2}\left(P^2+\frac{\kappa}{3}\rho a^{6}\right)=-\frac{a^2}{2}\left(P^2+\frac{\kappa}{3}\sum_{s=1}^n\rho_{0s}a^{3(1-\omega_s)}\right),
	\end{align}
	and write the Friedmann equation \eqref{eq_friedmann} as
	\begin{equation}
	\label{eq_conservation_V}
	\frac{\dot{a}^2}{2}+V(a)=0.
	\end{equation}
	This is the equation of conservation of energy for a particle $a=a(\tau)$, with unit mass and zero energy,
	evolving under the effective potential $V(a)$. Analyzing the form of this potential, one can then infer
	the global evolution of $a$. In particular, since its total energy is zero, the particle can only move along regions
	where the potential $V(a)$ is nonpositive, while the roots of the potential, where $\dot{a}$ vanishes,
	represent boundaries of such regions.
	Before commenting the general qualitative behavior of the solutions in the $\tau$ gauge, 
	let us first note that the vacuum and isotropic ($\rho=0=P)$ case is not dynamical,
	since the solution of the system is simply $a=a_0$. That is why, in some statements
	below, this degenerate $(\rho=0=P)$ case is excluded.
	
\begin{figure}
		\centering
		\begin{minipage}{0.46\textwidth}
			\centering
		\includegraphics[width=\linewidth]{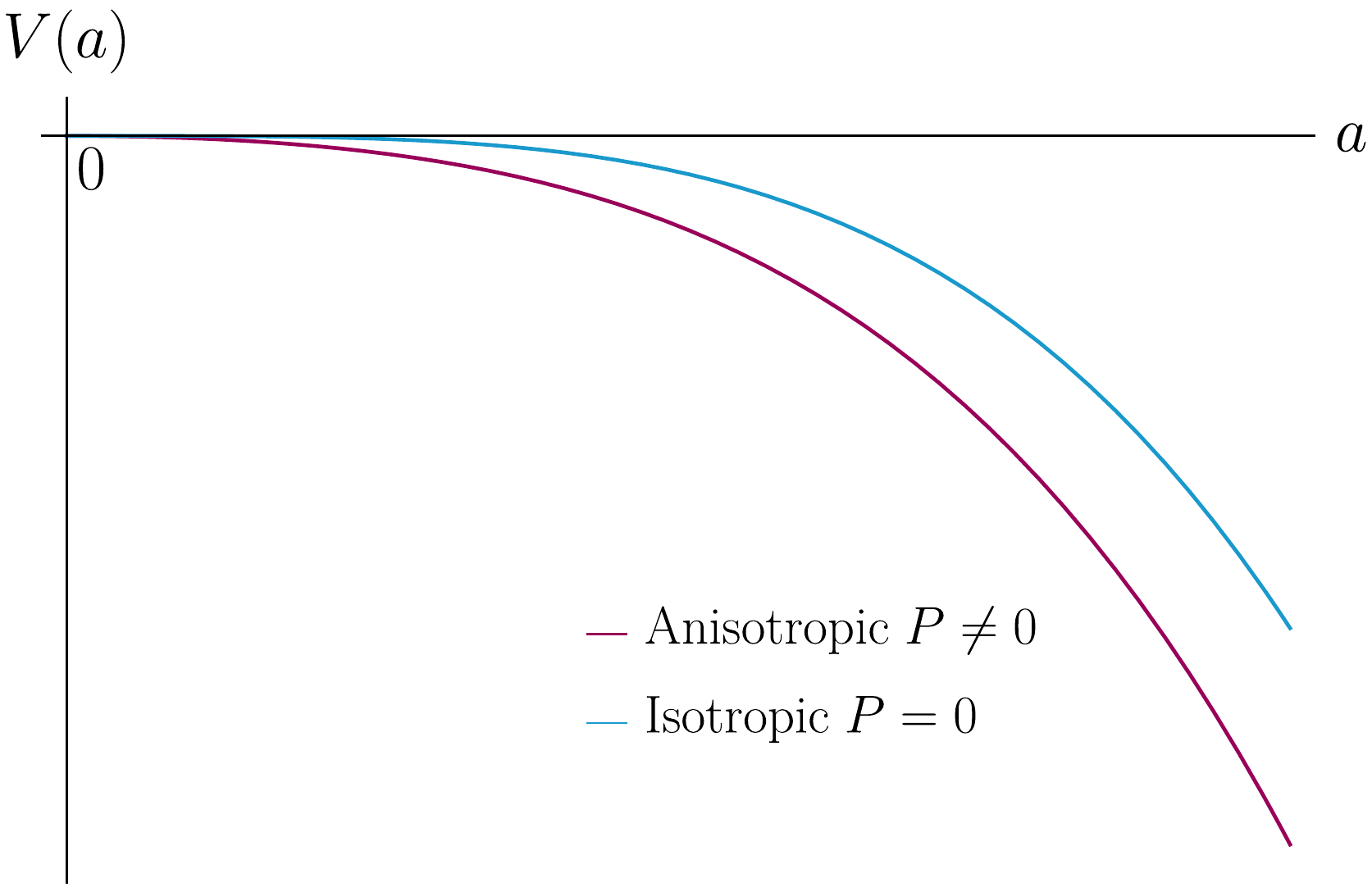}
		\caption{Shape of $V(a)$, as defined in \eqref{def_V_a}, for the specific scenario where all the densities $\rho_{0s}$ are positive or zero (excluding the case $\rho=0=P$), shown for both
		the isotropic $(P=0)$ and anisotropic $(P\neq 0)$ cases. As can be seen, in its domain $a>0$,
		$V(a)$ is a negative and monotonically decreasing function, with anisotropies ($P \neq 0$)
		making it more negative.
		}
		\label{fig:v_pos}
		\end{minipage}
		\hfill
		\begin{minipage}{0.46\textwidth}
			\centering
		\includegraphics[width=\linewidth]{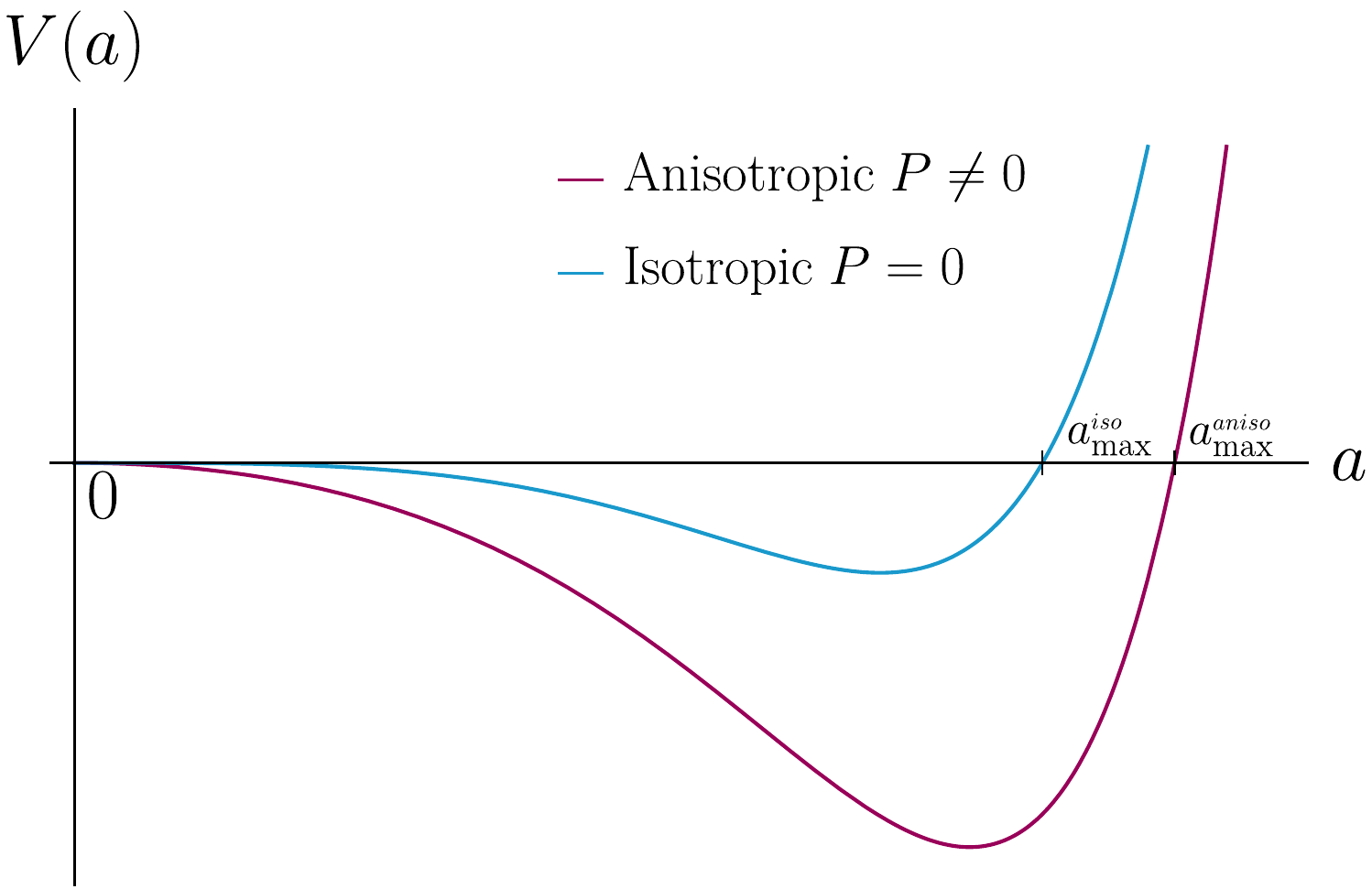}
		\caption{Shape of $V(a)$, as defined in \eqref{def_V_a}, for the specific scenario of a negative cosmological constant with the remaining densities $\rho_{0s}$ positive, shown for both the isotropic and anisotropic cases. As can be seen, $V(a)$ has a root in $a_{\max}$, which increases with the anisotropies ($P \neq 0$). Moreover, the latter also decrease the value of $V(a)$.
		}
		\label{fig:v_neg}
		\end{minipage}
	\end{figure}

	On the one hand, as shown in Fig.~\ref{fig:v_pos}, if all densities $\rho_{0s}$ are positive or zero (excluding the case $\rho=0=P$),
	for all $a>0$ the potential $V(a)$ takes negative values, and it is a monotonically decreasing function of $a$. At $a=0$, $V(a)$ vanishes and, for $P\neq 0$, it has a maximum there, while, for the isotropic case
	$P=0$ (with $\rho\neq 0$), $a=0$ is an inflection point. This implies that, in either case, $a=0$ is not a turning point, rather
	it is an unstable equilibrium point that the system will only reach in an infinite amount of time.\footnote{Note, however, that the amount of time is a gauge-dependent quantity. In a generic gauge with lapse $N$, the Friedmann equation
		\eqref{eq_conservation_V} reads $\frac{{a^{\prime}}^2}{2}+\frac{N^2}{a^6}V(a)=0$. For instance, for the cosmological time $t$,
		$N=1$, and the corresponding effective potential is $V(a)/a^6$, which, instead of a maximum, presents a divergence at $a=0$.
		Therefore, in cosmic time, the singularity is reached in finite time as $\frac{da}{dt}$ tends to infinity.}
	The image of the function $a=a(\tau)$ is the semi-infinite real line $(0,+\infty)$, and
	the velocity of the particle $\dot{a}$ increases with the value of $a$.
	Thus, choosing the outgoing (expanding) branch, the particle begins at $\tau\rightarrow-\infty$ 
	at the origin $a=0$ with $\dot{a}=0$, and monotonically increases its velocity $\dot{a}$ as it moves
	to larger values of $a$. 
	
		On the other hand, if some of the $\rho_{0s}$ are negative, the analysis is more involved since, in general,
	for $a>0$, the potential $V(a)$ will not be a monotonic function and it might have one (or several) roots.
	Such roots will bound allowed regions of $a$, and there might even exist more than
	one allowed region, defining different cosmological evolutions. In particular,
	below we present a detailed analysis for the case of two species with negative density,
	which, for certain ranges of the densities, describe two different cosmological evolutions.
	
	However, for this discussion, let us consider a simpler, though quite general, case:
	a model with negative cosmological constant $\Lambda<0$,
	for which $\omega_\Lambda=-1$ and $\rho_{0\Lambda}=\Lambda/\kappa<0$,
	while all the remaining species, if any, have $\rho_{0s}>0$
	and $\omega_s\in(-1,1)$. 
	We first note that, if $P=0$, the solution for a negative cosmological constant can only exist
	in the presence of other species so that the potential \eqref{def_V_a} is negative and thus \eqref{eq_conservation_V} can be satisfied. Therefore, the inconsistent case with $P=0$ and no other species than $\Lambda$ is excluded from
	this discussion. For all the other cases, the qualitative shape of the potential $V(a)$ is illustrated in Fig.~\ref{fig:v_neg},
	and one can show that it has a unique positive root,
	$a=a_{\rm max}$, where $V'(a_{\rm max})> 0$.\footnote{
		It is straightforward to prove and generalize this result for
		any model with one species with
		negative density $\rho_{0-}<0$ and barotropic index $\omega_-$,
		while the remaining species, if any, have $\rho_{0s}>0$ and larger barotropic index
		$\omega_s>\omega_-$. For such model,
		the potential $V(a)$ is a linear combination of powers of $a$,
		all with negative coefficients, except for the largest power $a^{5-3\omega_-}$,
		which is multiplied by the positive coefficient $\kappa |\rho_{0-}|/6$.
		Therefore, excluding the inconsistent case with $P= 0$ and no other species than $\rho_{0-}$,
		one can apply the Descartes rule for generalized polynomials (see, e.g., \cite{Jameson:2006}), and
		conclude that $V(a)$ has only one positive root $a_{\rm max}$. In addition, since $V(a)$ is negative in a neighborhood of $a=0$, and for $a\to +\infty$
		it tends to $V(a)\to\frac{\kappa}{6} |\rho_{0-}| a^{5-3\omega_-}$, which is positive, $V'(a_{\rm max})$ must be positive.
		}
	This represents a turning point, which is reached in
	a finite amount of time. If $P\neq 0$, the origin $a=0$ is a maximum and the potential $V(a)$ is negative in the domain $a\in(0,a_{\rm max})$. If $P=0$, the shape of the
		the potential around $a=0$ depends strongly on the specific $\omega_s$, though $V(a)$ is always negative in the region $a\in(0,a_{\rm max})$. Therefore, in all the cases that allow a solution with a negative cosmological constant, the evolution of the universe begins
	at the singularity $a=0$ at $\tau\to -\infty$, it expands until it reaches $a=a_{\rm max}$ at a finite
	value of $\tau$, where it recollapses, to tend again towards $a=0$ as $\tau\to +\infty$.

	In all of the cases, the effect of the anisotropies is to lower the value of $V(a)$ with respect to its
	isotropic ($P=0$) counterpart. According to \eqref{eq_conservation_V}, this implies  that, for any value of $a$, the velocity $|\dot{a}|$
	will be larger than in the isotropic case. Moreover, in solutions with a recollapse at $a=a_{\rm max}$, the value
	of $a_{\rm max}$ will increase as $P$ increases, so that the recollapse will happen at larger volumes.
	
	Finally, it is also interesting to analyze the qualitative behavior of the average Hubble factor \eqref{hubble},
	$$H=\frac{\dot{a}}{a^4}=\frac{{\rm sgn}(\dot{a})}{a^4}\sqrt{-2 V(a)}.$$
	From this expression, it is easy to check that it diverges
	towards the singularity $a\to 0$ in all the cases, except for the case with $P=0$, $\Lambda>0$,
	and no other matter content. In the latter case, $H\to{\rm sgn}(\dot{a})\sqrt{\Lambda/3}$ as $a\to 0$.
	Concerning large volumes, that is, as $a\to+\infty$, on the one hand, for monotonic solutions with $\Lambda\geq 0$,
	the Hubble factor tends to ${\rm sgn}(\dot{a})\sqrt{\Lambda/3}$.
	On the other hand, for recollapsing solutions
	(with $\Lambda<0$, or some other exotic species with a negative density that triggers the recollapse),
	the image of $H(\tau)$
	is the whole real line, and it vanishes at $a_{\rm max}$. In general, for a given value of $a$,
	the anisotropies increase the value of $|H|$ as compared to its isotropic counterpart.
	However, the effect of the anisotropies in $H$ dies off as $a$ expands, and under the presence of other species in the model with $\omega_s<1$, they will be subdominant for large volumes. 
	
	\section{Explicit particular solutions}
	\label{sec.solutions}
	Now that we have described the qualitative behavior of the system in the different cases,
	we will obtain certain explicit solutions for relevant scenarios. It is clear that
	the Friedmann equation \eqref{eq_friedmann} can be reduced to an integral,
	\begin{equation}\label{integral}
	\int \frac{da}{\sqrt{- 2 V(a)}} = \pm (\tau-c),
	\end{equation}
	with an integration constant $c$ and the global sign $\pm$ corresponding to the sign of $\dot{a}$.
	Generically this integral cannot be computed explicitly, except in a number of cases.
	Let us therefore detail certain instances where it can be explicitly performed.
	Note that $c$ and the global sign are just symmetries of the solution:
	given a solution $a=f(\tau)$, $f(-\tau)$ is also a solution, which describes the same dynamics though backwards in time,
	as well as $f(\tau-c)$, which just shifts the origin of time. For compactness, we will
	appropriately choose a convenient $c$ in each case, while, for monotonic solutions,
	the positive global sign will be taken so that $\dot a>0$, and the universe is expanding.
	We recall also that the stiff-matter component is not explicitly considered, though one could add it
	to any of the solutions below by simply performing the replacement $P\rightarrow (P+\frac{\kappa}{3}\rho_{0{\rm stiff}})^{1/2}$.
	
		\subsection*{Vacuum}
		
		The simplest case corresponds to vacuum: $\rho=0$. In such case, the above integrand does not depend on $a$,
		and one obtains
		\begin{equation}
		\label{sol_a_Kasner_vacuum}
		a=a_0 e^{P\tau}.
		\end{equation}
		This is the well-known Kasner solution \cite{Kasner:1921zz}.
		In order to see its usual form as a power law in
		terms of the cosmological time, one can perform the change $t\to \tau$ by solving Eq.~\eqref{eq_cosmological_time}, that is,
		\begin{align}\label{cosmological_time_vacuum}
		t=\frac{a_0^3}{3P}e^{3P\tau},
		\end{align}
		and obtain
		\begin{equation}
		a\propto t^{1/3}.
		\end{equation}
		Moreover, by inverting the definition \eqref{def_misner_var}, and taking into account the evolution \eqref{sol_beta_+_Kasner} and \eqref{sol_beta_-_Kasner} of $\beta_{\pm}$, we can explicitly obtain the evolution of the different scale factors $a_i$ in terms of the cosmological time,
		\begin{align}\label{evol_Kasner}
		a_1\propto t^{p_1},\quad
		a_2\propto t^{p_2},\quad
		a_3\propto t^{p_3},\quad
		\end{align}
		where $p_i$ are the so-called Kasner exponents,
		\begin{align}\label{def_Kasner_exp}
		\begin{aligned}
		p_1&:=\frac{1}{3P}
		\left(
		P+\sqrt{3}p_-+p_+
		\right),
		\quad
		p_2&:=\frac{1}{3P}
		\left(
		P-\sqrt{3}p_-+p_+
		\right),
		\quad
		p_3&:=\frac{1}{3P}
		\left(
		P-2p_+
		\right).
		\end{aligned}
		\end{align}
		Since $P=(p_+^2+p_-^2)^{1/2}$, it is easy to verify that these exponents obey the usual relations,
		\begin{align}\label{Kasner_exp_prop}
		p_1+p_2+p_3=p_1^2+p_2^2+p_3^2=1,
		\end{align}
which implies that at least one of the Kasner exponents $p_i$ is nonpositive,
		and thus its corresponding scale factor either expands or remains constant as the universe tends towards the singularity.
		
		It is important to note that for a stiff-matter content, since we have to perform the change $P\rightarrow (P+\frac{\kappa}{3}\rho_{0{\rm stiff}})^{1/2}$, this property is modified, namely,
		\begin{align}\label{Kasner_exp_prop_stiff}
		p_1+p_2+p_3=1\quad\text{and}\quad p_1^2+p_2^2+p_3^2=\frac{1}{3}\left(1+\frac{2}{1+\frac{\kappa}{3P^2}\rho_{0\rm stiff}}\right),
		\end{align}
		which, provided $\rho_{0\rm stiff}>0$, allows all three exponents $p_i$ to be positive for a range of values of $p_{\pm}$.	

		\subsection*{A single species}	

\begin{figure}
			\centering
			\begin{minipage}{0.46\textwidth}
				\centering
				\includegraphics[width=\linewidth]{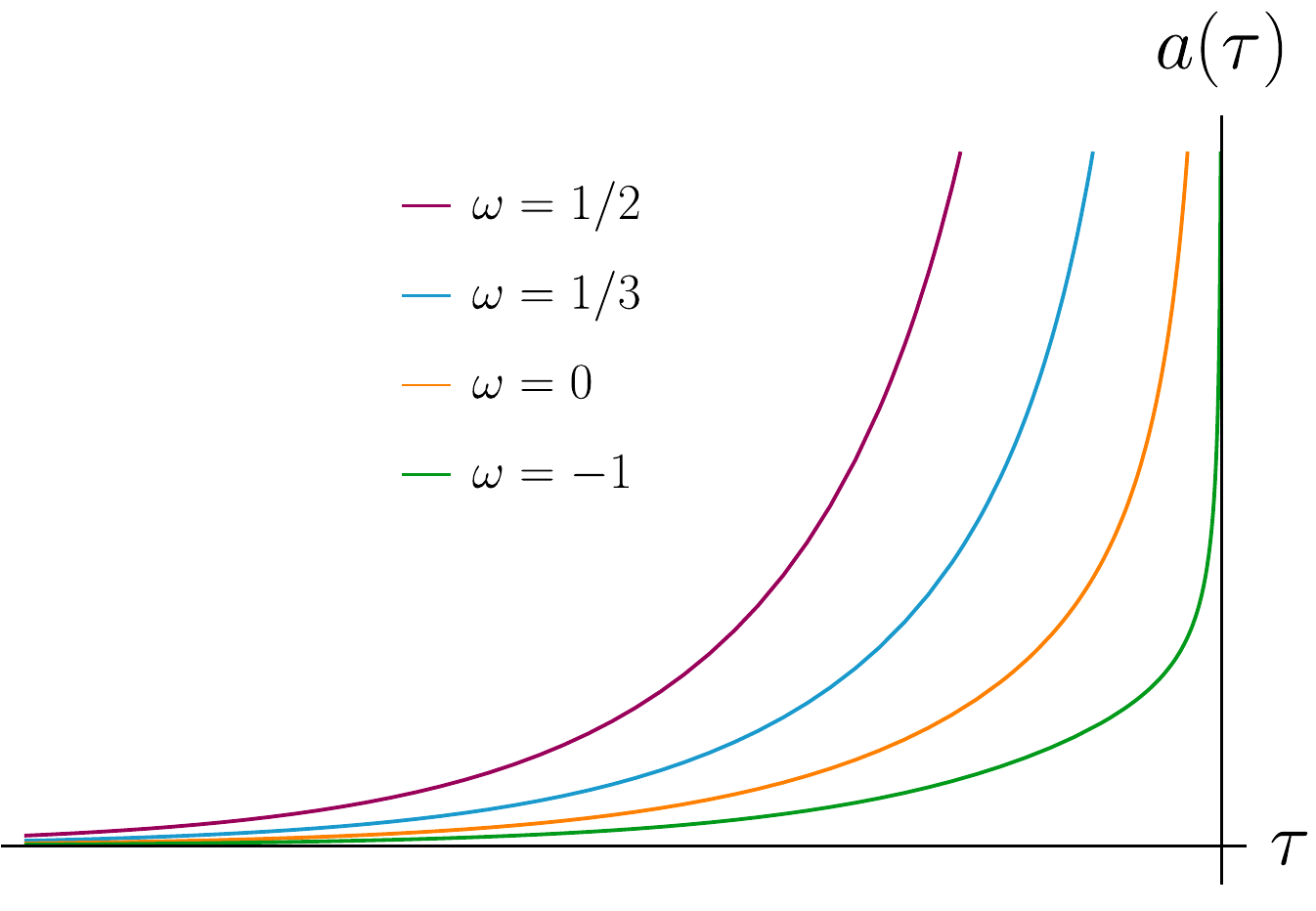}
				\caption{Evolution of the average scale factor $a$ in terms of $\tau$ given by \eqref{sol_a_single_matter_pos}, for a single species and a positive density $\rho_{0}>0$. Different colors correspond to different barotropic indices $\omega$. As can be seen, for a given $\tau$, a larger value of $\omega$ implies a larger value of $a$.
				}
				\label{fig:a_pos}
			\end{minipage}
			\hfill
			\begin{minipage}{0.46\textwidth}
				\centering
				\includegraphics[width=\linewidth]{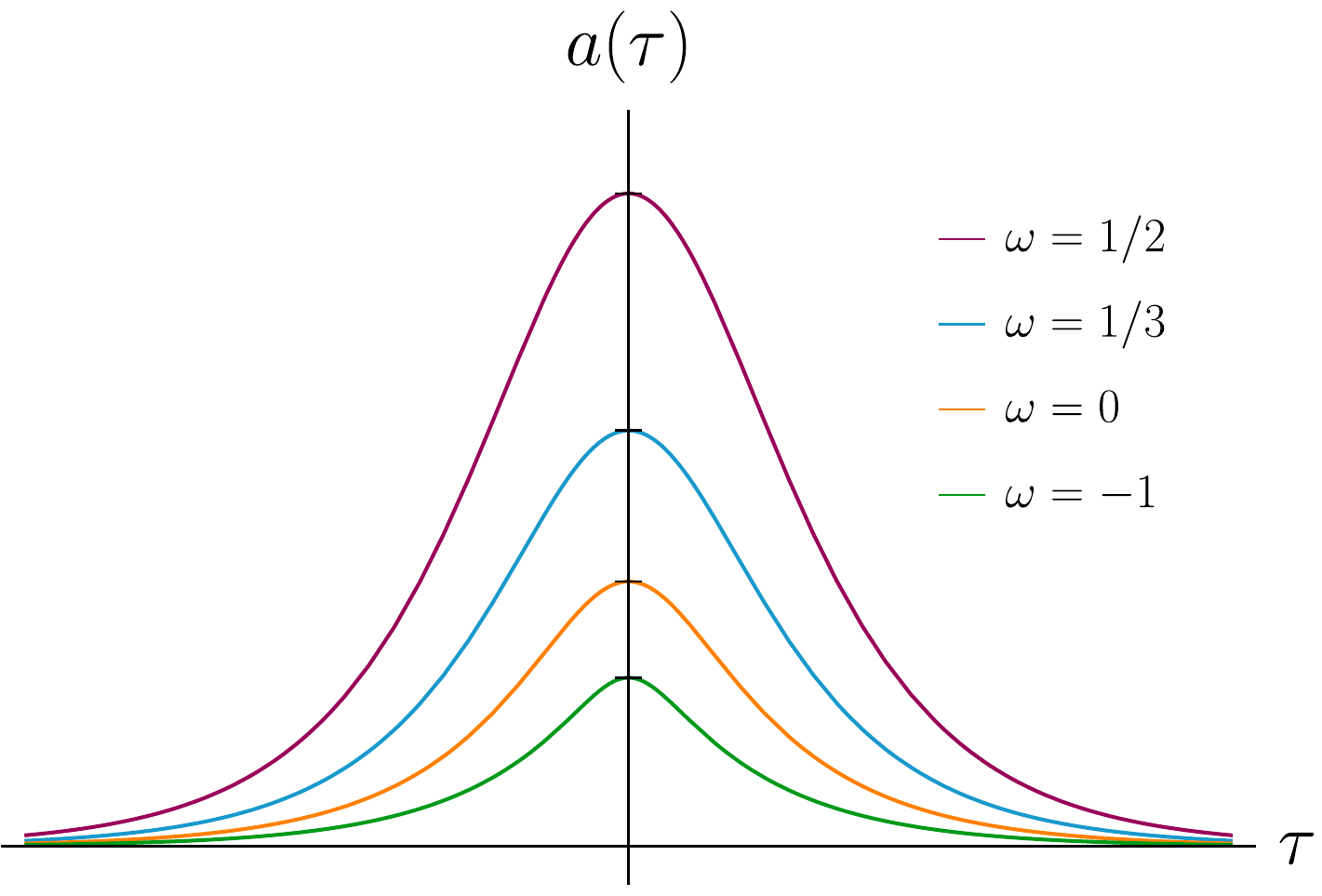}
				\caption{Evolution of the average scale factor $a$ in terms of $\tau$ given by \eqref{sol_a_single_matter_neg}, for a single species and a negative density $\rho_{0}<0$. Different colors correspond to different barotropic indices $\omega$. As can be seen, for larger $\omega$, the $a$ is larger at each value of $\tau$.
				Hence, the corresponding maximum value $a_{\max}$, given by \eqref{a_max},
				 is also larger.}
				\label{fig:a_neg}
			\end{minipage}
		\end{figure}

		Let us consider now the presence of only one species
		with a generic barotropic index $\omega\neq 1$ and $\rho_{0s}=\rho_0$. In this case, considering a
		positive density $\rho_0>0$, the solution reads as follows,:
		\begin{align}\label{sol_a_single_matter_pos}
		a=\left[\frac{3P^2}{\kappa\rho_0\sinh^2\left({\frac{3}{2}(1-\omega)P \tau}\right)}\right]^{\frac{1}{3(1-\omega)}},
		\end{align}
		with domain $\tau\in(-\infty,0)$. The form of this function is shown in Fig. \ref{fig:a_pos} for different values of $\omega$. The isotropic
		case is included in the above expression as the limit $P\to0$, and it explicitly reads
		\begin{equation}
		\label{limit_isotropic}
		a\propto|\tau|^{-\frac{2}{3(1-\omega)}}.
		\end{equation}		
		This solution contains, as a particular case, the universe filled with dust $(\omega=0)$, which
		corresponds to the well-known Heckmann-Sch\"ucking \cite{Heckmann-Schucking} solution.
		Another particular case is the one presented in Ref.~\cite{deCesare:2019suk}, involving an ekpyrotic fluid ($\omega=3$) with an additional 	stiff-matter component, which can be included in \eqref{sol_a_single_matter_pos} by the replacement commented above $P\to (P+\frac{\kappa}{3}\rho_{0{\rm stiff}})^{1/2}$.
		
		The solution \eqref{sol_a_single_matter_pos} has been found in several (implicit) forms in the literature (see, for instance, Ref.~\cite{Jacobs:1968A}),
		but, to the best of our knowledge, it has not been given in this simple and explicit form, which we could
		obtain due to the gauge choice. Although the change to the cosmological time cannot be carried out explicitly
		for finite values of $\tau$, as will be detailed in the next section, where we will study the asymptotics of this
		solution, it is possible to do it for the limit of small and large volumes.
		
		Moreover, let us also consider a negative density $\rho_0<0$, which includes the particular case of a negative cosmological constant
		and no other matter content.
		In such a case, as commented above, there is no solution for the isotropic $(P=0)$ model, while, for $P\neq 0$,
		the scale factor evolves as
		\begin{align}\label{sol_a_single_matter_neg}
		a=\left[\frac{3P^2}{\kappa|\rho_0|\cosh^2\left({\frac{3}{2}(1-\omega)P \tau}\right)}\right]^{\frac{1}{3(1-\omega)}},
		\end{align}
		with domain $\tau\in(-\infty,\infty)$. As can be seen in Fig. \ref{fig:a_neg}, this solution contains an initial and a final singularity at
		$\tau\to-\infty$ and $\tau\to+\infty$, respectively, while $\tau=0$ corresponds to a recollapse.
		The domain of $a$ is
		thus bounded to $a\in(0,a_{\rm max}]$, where
		\begin{align}\label{a_max}
		a_{\max}:=
		\left(
		\frac{
			3P^2}{\kappa |\rho_0|}\right)^{\frac{1}{3(1-\omega)}}.
		\end{align}
		
		\subsection*{Two species}
		
				If there are two species $n=2$, and their corresponding barotropic indices
		$\omega_1$ and $\omega_2$ obey $(1-\omega_2)=2(1-\omega_1)$, the potential $V(a)$ is a simple polynomial of $a$; then,
		the integral \eqref{integral} can be computed explicitly and it has a particularly simple form.
		Note that, since $\omega_2=2\omega_1-1$ and we are assuming that both $\omega_1$ and $\omega_2$ lay in the range $[-1,1)$, necessarily $\omega_1\in[0,1)$. Consequently, $\omega_2<\omega_1$, and we will thus use the names $\omega_-:=\omega_2$ and $\omega_+:=\omega_1$ for the barotropic indices, and $\rho_{0-}:=\rho_{02}$ and $\rho_{0+}:=\rho_{01}$ for
		the densities.

		This case is a generalization of the models analyzed in Refs. \cite{Khalatnikov:2003ph,Kamenshchik:2009dt},
		where they considered a Bianchi I universe filled with stiff matter, dust and a cosmological constant $\Lambda$
		(both positive and negative). As will be detailed below,
		these cases can be recovered from our results by simply imposing $\omega_-=-1$, $\omega_+=0$, $\rho_{0+}=\rho_{0{\rm dust}}$,
		$\rho_{0-}=\Lambda/\kappa$, and including the stiff-matter contribution by the replacement $P\to (P^2+\frac{\kappa}{3}\rho_{0{\rm stiff}})^{1/2}$. There are, however, other cases
		that may be of interest and are included in this analysis, for instance,
		a Bianchi I universe filled with radiation $(\omega_+=1/3)$
		and cosmic strings $(\omega_-=-1/3)$.

		For completeness, below we will present the explicit solution for all the possible signs of the densities. But, let us
		first comment several properties that can be inferred from the analysis of the potential $V(a)$.
		In particular, note that for small scales $a\to 0$, $\rho_{0+}$ will be dominant, while for large scales $a\to +\infty$, $\rho_{0+}$ will be negligible and $\rho_{0-}$ will dominate. If both $\rho_{0+}$ and $\rho_{0-}$ are positive, the potential $V(a)$ does not have any positive root,
		it is negative all along $a\in [0,+\infty)$, and the model describes an indefinite cosmological expansion (see Fig.~\ref{fig:v_2_pos}). But
		negative values of the densities $\rho_{0+}$ and $\rho_{0-}$ may introduce positive roots in $V(a)$, which, as commented
		above, are turning points that bound $a$. More precisely, $\rho_{0-}$ being the dominant species for $a\to +\infty$,
		if $\rho_{0-}<0$, this negative energy will make $V(a)$ to take positive values at large values of $a$, independently of the value of $\rho_{0+}$.  Thus, for $\rho_{0-}<0$, the potential $V(a)$ will have one, and only one,
			positive root $a_{\rm max}$. This value will bound $a$ from above, triggering a recollapse
			(see Fig.~\ref{fig:v_2_rho_-_neg}).
			
		However, the case $\rho_{0+}<0$ and $\rho_{0-}>0$ is more involved and the existence
		of turning points depends on whether $\rho_{0-}$ exceeds certain threshold $\rho_{\rm threshold}:=\kappa \rho_{0+}^2/12P^2$. More precisely, if
		$\rho_{\rm threshold}<\rho_{0-}$, there are no real roots of $V(a)$ and one gets an indefinite cosmological expansion.
		However, if the second species is not so energetic and $0\leq\rho_{0-}<\rho_{\rm threshold}$, $V(a)$
		has two positive roots $a_\pm$, which leads to two different and independent
		cosmological evolutions. In one of them $a$ is bounded to $a\in(0,a_-]$ and it describes a recollapsing cosmology with an initial and a final singularity. In the other one, $a\in[a_+,+\infty)$ is bounded from below, and it provides a
		singularity-free cosmology that begins with infinite volume, reaches a minimum at $a_+$,
		where it bounces back to expand then forever. These two different scenarios are illustrated in Figs.~\ref{fig:v_2_rho_-_pos_u_thres} and \ref{fig:v_2_rho_-_pos_d_thres}, respectively.

		Let us now detail the explicit form of the solution for the different cases. On the one hand,
		for {{$\rho_{0+}>0$}}, and any sign of {{$\rho_{0-}$}}, there is only one solution and it reads
		\begin{align}
		\label{sol_bianchi_I_k=2_pos}
		a&=\left[
		{\frac{\kappa\rho_{0+}}{3P^2}\sinh^2\left({\frac{3}{2}(1-\omega_+)P\tau}\right)}
		-\frac{\rho_{0-}}{\rho_{0+}}e^{3 (1-\omega_+) P \tau}
		\right]^{-\frac{1}{3(1-\omega_+)}}.
		\end{align}
		This evolution can be seen in Fig.~\ref{fig:evol_a_rho01_pos}.
		On the other hand, for $\rho_{0+}<0$, any value of $\rho_{0-}$, and $P\neq 0$, there is a solution that can be written as
		\begin{align}
		\label{sol_bianchi_I_k=2_neg}
			a&=\left[
		{\frac{\kappa|\rho_{0+}|}{3P^2}\cosh^2\left({\frac{3}{2}(1-\omega_+)P \tau}\right)}
		-\frac{\rho_{0-}}{|\rho_{0+}|}e^{3(1-\omega_+) P \tau}
		\right]^{-\frac{1}{3(1-\omega_+)}},
		\end{align}
		which is depicted in Fig.~\ref{fig:evol_a_rho01_neg_gen}.
		However, if $\rho_{0+}<0$ and $\rho_{0-}\in[0, \rho_{\rm threshold}]$, with $\rho_{\rm threshold}=\frac{\kappa \rho_{0+}^2}{12P^2}$, there is an additional solution of the form
		\begin{align}
		\label{sol_bianchi_I_k=2_neg_special}
			a&=\left[-
		{\frac{\kappa|\rho_{0+}|}{3P^2}\sinh^2\left({\frac{3}{2}(1-\omega_+)P\tau}\right)}
		+\frac{\rho_{0-}}{|\rho_{0+}|}e^{3 (1-\omega_+) P \tau}
		\right]^{-\frac{1}{3(1-\omega_+)}},
		\end{align}
		which is shown in Fig.~\ref{fig:evol_a_rho01_neg_special}.
		In the solutions \eqref{sol_bianchi_I_k=2_pos} and \eqref{sol_bianchi_I_k=2_neg_special}
		the isotropic case is included as the limit $P\to 0$, but \eqref{sol_bianchi_I_k=2_neg}
		is not defined in such limit as this solution does not exist for $P=0$.
		Note that, if $P=0$ then $\rho_\text{threshold}\to +\infty$, and thus  \eqref{sol_bianchi_I_k=2_neg_special} is the only solution for $\rho_{0+}<0$, provided that $\rho_{0-}$ is positive.

		As commented above, the qualitative behavior of these functions, and the corresponding range of $a$,
 is determined by the values of $P$, $\rho_{0+}$, and $\rho_{0-}$. Below we summarize the different cases.

\begin{itemize}
			\item For cases with $\{0<\rho_{0+}$ and $0<\rho_{0-}\}$ or $\{\rho_{0+}<0$ and $\rho_{\rm threshold}<\rho_{0-}\}$,
			as can be seen in Figs.~\ref{fig:v_2_pos} and \ref{fig:v_2_rho_-_pos_u_thres},
			the potential has no positive roots, and the evolution
			of the scale factor is given by \eqref{sol_bianchi_I_k=2_pos} or \eqref{sol_bianchi_I_k=2_neg}, respectively.
These solutions describe an indefinite expansion $a\in(0,+\infty)$, with the domain $\tau\in(-\infty,\tau_0)$, where
			\begin{align*}
			\tau_0:=-\frac{1}{3P(1-\omega_+)}\ln\left(
			{\rm sgn}(\rho_{0+})+\frac{2\sqrt{3}P\sqrt{\rho_{0-}}}{|\rho_{0+}|\sqrt{\kappa}}
			\right)
			\end{align*}
			is the time when infinite volume is reached,
			while the singularity is located at $\tau\to -\infty$.
			A particular case of this solution corresponds to the case analyzed in Ref.~\cite{Khalatnikov:2003ph},
			with dust ($\omega_+=0$ and $\rho_{0+}=\rho_{0{\rm dust}}$), a positive cosmological constant ($\omega_-=-1$ and $\rho_{0-}=\Lambda/\kappa$), and stiff matter [included as $P\to(P^2+\frac{\kappa}{3}\rho_{0{\rm stiff}})^{1/2}$].

\item For cases with $\rho_{0-}<0$ and $P\neq 0$, the potential $V(a)$ has a single positive root (see Fig.~\ref{fig:v_2_rho_-_neg}), the solution is  given
either by \eqref{sol_bianchi_I_k=2_pos} if $\rho_{0+}>0$, or by \eqref{sol_bianchi_I_k=2_neg} if $\rho_{0+}<0$,
		with domain $\tau\in\mathbb{R}$.
The range of $a$ is $a\in(0,a_{\max}]$, and thus there is a recollapse at
	$$a_{\max}=\left[\frac{1}{2|\rho_{0-}|}\left(
			\rho_{0+}+\sqrt{\rho_{0+}^2+\frac{12P^2|\rho_{0-}|}{\kappa}}
			\right)\right]^{\frac{1}{3(1-\omega_+)}},$$
		and two singularities with $a=0$. The initial one corresponds to $\tau\to -\infty$, and the final one, after the recollapse, to $\tau\to +\infty$.
		From this general solution, one can reproduce the case
		analyzed in Ref.~\cite{Kamenshchik:2009dt}, where they consider dust,
		a negative cosmological constant $\Lambda<0$, and stiff matter, by simply imposing $\omega_+=0$,
		$\rho_{0+}=\rho_{0{\rm dust}}$,
		$\omega_-=-1$, $\rho_{0-}=\Lambda/\kappa$,  and $P\to(P^2+\frac{\kappa}{3}\rho_{0{\rm stiff}})^{1/2}$. 
		
		\item For cases with $\{\rho_{0+}<0$ and $0\leq\rho_{0-}\leq \rho_{\rm threshold}\}$ the potential $V(a)$ has two positive roots (see Fig.~\ref{fig:v_2_rho_-_pos_d_thres}),
		\begin{equation}
		a_\pm:=\left[\frac{1}{2\rho_{0-}}\left(
			|\rho_{0+}|\pm\sqrt{\rho_{0+}^2-\frac{12P^2\rho_{0-}}{\kappa}}
			\right)\right]^{\frac{1}{3(1-\omega_+)}},
		\end{equation}
		and, as explained above, this case describes two different and independent cosmologies.

	On the one hand, it describes a finite universe with an evolution given by the functional form \eqref{sol_bianchi_I_k=2_neg}, with image
		$a\in(0,a_-]$, which thus
		presents an initial and a final singularity, as well as a recollapse at $a_-$, as can be seen in Fig.~\ref{fig:evol_a_rho01_neg_gen}.
			On the other hand, it also describes a singularity-free universe, with an evolution given by the functional form \eqref{sol_bianchi_I_k=2_neg_special} and depicted in Fig.~\ref{fig:evol_a_rho01_neg_special},
			which leads to an image $a\in[a_+,+\infty)$, and thus undergoes a bounce at $a=a_+$.
			An interesting property of this cosmology is that, even if one needs to assume an exotic fluid with $\rho_{0+}<0$
			to resolve the singularity, for large volumes $a\to +\infty$ the density of such fluid decays until becoming negligible, and the other nonexotic component $\rho_->0$ dominates completely the evolution.
			 Let us also mention that, in the isotropic case $P\to 0$, the threshold energy $\rho_{\rm threshold}$
			tends to infinity and, thus, the exotic $\rho_{0+}$ species is able to produce a bounce and provide a
			singularity-free cosmological evolution, independently of the energy contained in the $\rho_{0-}$ component.
			
	In the degenerate case with $\rho_{0-}=\rho_{\rm threshold}$, both roots coincide $a_+=a_-$, $V(a)$ presents a maximum there, and thus $a=a_+=a_-$ is an unstable equilibrium point. This implies that, unlike in the general case that the system reaches $a_\pm$ in finite time, in the degenerate case $a_\pm$ will be reached	only in an infinite amount of time. The time evolution for this case is shown in Fig.~\ref{fig:evol_a_rho01_deg}.
\end{itemize}
		\subsection*{Several species}
		
		Finally, for the case of several species such that all $(1-\omega_s)$ are proportional to $(1-\omega_1)$,
	more specifically, $(1-\omega_s)=s(1-\omega_1)$,
		the integral can be performed in terms of elliptic functions, but its form is extremely complicated
		and we will refrain from writing it explicitly.

	\section{Asymptotic behavior}
	\label{sec.asymptoticsBI}
	Let us now study the asymptotic behavior, both towards the singularity and towards large volumes,
	of the Bianchi I model filled with several barotropic fluids. As commented in Subsec.~\ref{sec:equation_of_state}, due to their different
	barotropic indices, the scaling of the species differs. For $a\to0$, the fluid with the maximum
	barotropic index $\omega_+$ will dominate, while for $a\to+\infty$ all the species will be
	negligible except the one with the minimum $\omega_-$. Therefore, we will consider the regime where only
	one fluid is dominant, so we can then use the explicit solution \eqref{sol_a_single_matter_pos}.
	Since recollapsing or bouncing cosmologies with exotic matter fields
	 may not reach the limit of either large or small volumes, for simplicity, here we will assume that in both limits the corresponding dominant species has a positive energy density $\rho_0$.

	 As commented in the previous section, we will analyze the curvature by means of the Kretschmann scalar \eqref{def_Kretschmann} and the Weyl square
	\eqref{Weylsquare}, which are related by Eq.~\eqref{W2K}. Specifically, considering a universe with a single species with $\rho_0>0$ and $\omega\in[-1,1)$, which follows
	the evolution \eqref{sol_a_single_matter_pos}, for $P\neq 0$ the Kretschmann scalar reads
	\begin{align}
	\nonumber
	K=&\frac{K_0}{P^3}
	\left|\sinh\left(
	\frac{3}{2}P(1-\omega)\tau
	\right)\right|^{\frac{4(1+\omega)}{1-\omega}}
	\bigg[
	4P^3
	\cosh\left(
	6P(1-\omega)\tau
	\right)
	-8P^3\cosh\left(
	3P(1-\omega)\tau
	\right)
	\\
	&
	-16p_+(3p_-^2-p_+^2)
	\sinh\left(
	3P(1-\omega)\tau
	\right)
	\sinh^2\left(
	\frac{3}{2}P(1-\omega)\tau
	\right)
	+3P^3(3+2\omega+3\omega^2)
	\bigg],
	\label{kretschmann}
	\end{align}
	where $K_0:=3^{-(3+\omega)/(1+\omega)}P^4\left(\kappa\rho_0/
	P^2
	\right)^{4/(1-\omega)}$, while for the isotropic case $P=0$,
	\begin{align}\label{kretschmann_isotropic}
		K\propto |\tau| ^{\frac{4(1+\omega)}{1-\omega}}.
	\end{align}
From here, it can be noted that, when $\omega=-1$, the Kretschmann scalar is a constant, more specifically, $K=8\kappa^2\rho_0^2/3$. Moreover, for vacuum $\rho_0=0$, the Kretschmann scalar reads
	\begin{align}\label{Kretschmann_vacuum}
		K=96e^{-12P\tau}P(P-2p_+)^2(P+p_+),
	\end{align}
and we recall that for this latter case $P$ cannot be vanishing. This expression also includes the stiff-matter case ($\omega=1$), by simply performing the change
$P\to(P^2+\frac{\kappa}{3}\rho_{0\rm{stiff}})^{1/2}$. Hence, for vacuum and stiff-matter content the scaling of $K$ can straightforwardly be seen in \eqref{Kretschmann_vacuum} for both $\tau\to\pm\infty$.

\subsection{Isotropization towards large volumes}

	On the one hand, according to \eqref{sol_a_single_matter_pos}, the limit towards big volumes, $a\to +\infty$, corresponds to $\tau\to 0$. Hence, in this regime,
	the scale factor goes as a negative power in $\tau$,
	\begin{align}\label{aisotropized}
	a&\approx a_{0}|\tau|^{-\frac{2}{3(1-\omega)}},
	\end{align}
	where $a_0:=\left[
	\frac{3}{4}\kappa\rho_0(1-\omega)^2 \right]^{-\frac{1}{3(1-\omega)}}$,
	and one should take into account that the evolution \eqref{sol_a_single_matter_pos} is only valid for $\omega\in[-1,1)$ and $\rho_0>0$, and thus also this approximation. Moreover, taking this limit in the evolution \eqref{sol_beta_+_Kasner} and \eqref{sol_beta_-_Kasner} of the shape parameters leads to
	\begin{align}\label{evolution_Kasner_big_volumes}
	\beta_+&=k_+,
	\\
	\beta_-&=k_-.
\end{align}
	Therefore, for large volumes, the shape parameters tend to a constant value.
	However, as explained above, the constants $k_\pm$
	can be reabsorbed in the coordinates; hence, we can set $\beta_{\pm}=0$, and thus this corresponds to an isotropic universe.
	In fact, \eqref{aisotropized} is identical to the isotropic solution \eqref{limit_isotropic}. Consequently, the system isotropizes at big volumes, tending to a flat FLRW universe. In order to see its evolution in
	terms of the cosmological time, one just needs to perform the integral \eqref{eq_cosmological_time}, which provides, for $\omega\in(-1,1)$, 
	\begin{align}\label{cosmological_time_FLRW}
	&t\propto |\tau|^{-\frac{1+\omega}{1-\omega}},
	\end{align}
while, for $\omega=-1$,
	\begin{align}\label{cosmological_time_FLRW_w=-1}
		t=-\frac{\ln |\tau|}{\sqrt{3\kappa\rho_0}}.
	\end{align}
	These relations can be used in \eqref{aisotropized} to write, for $\omega\in(-1,1)$,
	\begin{align}
	\label{aisotropized_cosmological}
	a\propto t^{\frac{2}{3(1+\omega)}},
	\end{align}
	and, for $\omega=-1$,
	\begin{align}\label{aisotropized_cosmological_w=-1}
		a\propto e^{t\sqrt{\frac{\kappa\rho_0}{3}}},
	\end{align}
	which is the usual behavior of the scale factor for a flat FLRW universe.

	Concerning the curvature,
	in this limit of large volumes, the Kretschmann scalar \eqref{kretschmann} scales as in the isotropic case \eqref{kretschmann_isotropic},
	\begin{align}\label{kretschmann_big_volumes}
	K\propto |\tau| ^{\frac{4(1+\omega)}{1-\omega}}.
	\end{align}
In cosmological time \eqref{cosmological_time_FLRW}, for $\omega\in(-1,1)$, it reads as
	\begin{align}\label{kretschmann_big_volumes_tau}
	K\propto t^{-4},
	\end{align}
while, for $\omega=-1$, it is just a constant, $K=8\kappa^2\rho_0^2/3$ specifically, as in the isotropic case.
	Moreover, making use of \eqref{aisotropized_cosmological},
	it is also interesting to write it in terms of $a$,
	\begin{align}
\label{kretschmann_big_volumes_a}
	K\propto a^{-6(1+\omega)},
	\end{align}
	valid again for $\omega\in(-1,1)$.
	Therefore, for large volumes $a\to +\infty$, the Kretschmann scalar vanishes for $\omega\in(-1,1)$.
	Finally, from relation \eqref{Weyl_large_vol} it is straightforward to compute
	the scaling of the Weyl square,
	\begin{align}
		\label{weyl_square_big_volumes}
		W^2\propto a^{-6(1+\omega)},
	\end{align}
	which turns out to be identical to the scaling of the Kretschmann scalar, and thus it also vanishes at large volumes if $\omega\in(-1,1)$. However, if $\omega=-1$, the Kretschmann scalar is a constant, $K=8\kappa^2\rho_0^2/3$, and, according to \eqref{Weyl_large_vol}, the Weyl square exactly vanishes: $W^2=0$.

	\subsection{Asymptotics towards the singularity and blowup of the curvature}

	On the other hand, the limit $a\to 0$ corresponds to the system approaching the singularity, which, for the expanding branch, takes place at $\tau\to -\infty$. Therefore, by taking this limit in \eqref{sol_a_single_matter_pos}, the evolution of $a$ can be approximated as
	\begin{align}\label{evolution_alpha_Kasner_singularity}
	a&\approx\left(\frac{12P^2}{\kappa\rho_0}\right)^{\frac{1}{3(1-\omega)}}e^{P\tau},
	\end{align}
	for $P\neq 0$ and $\omega\neq 1$. For the isotropic ($P=0$) case with $\omega\neq 1$, the evolution is given by \eqref{limit_isotropic}.
	Hence, unless $P=0$, in this limit $a$ evolves exponentially in $\tau$, mirroring precisely the evolution of the vacuum (Kasner) solution, as given in Eq.~\eqref{sol_a_Kasner_vacuum}. In order to translate this to the cosmological time $t$, we can just consider the relation obtained for the vacuum solution \eqref{cosmological_time_vacuum}, which leads to the form
	\begin{align}\label{sol_cosmological_time_sing_limit}
	a\propto t^{\frac{1}{3}}.
	\end{align}

	To finish with this analysis, let us evaluate the Kretschmann scalar \eqref{kretschmann} in this limit. For the $P=0$ case, it scales as a power law \eqref{kretschmann_isotropic}, and therefore always diverges towards the singularity, regardless of the matter type. For $\rho_0>0$, $\omega\neq 1$, and $P\neq 0$ in the limit towards the singularity, $\tau\to -\infty$,
	\begin{align}\label{kretschmann_singularity}
	K\approx\frac{K_0}{4^{(1+3\omega)/(1-\omega)}}\,e^{-12P\tau}\,\sin^2\left(\frac{3\theta}{2}
	\right),
	\end{align}
	where we have defined $\cos\theta:=-p_+/P$ and $\sin\theta:=- p_-/P$. This scaling is identical to the behavior of the Kretschmann scalar in the
	exact vacuum model \eqref{Kretschmann_vacuum}.
Again, making use of the relations \eqref{cosmological_time_vacuum} and \eqref{evolution_alpha_Kasner_singularity}, we can write
this result in the comoving gauge,
	\begin{align}
	\label{kretschmann_singularity_cosmological}
	K\approx\frac{64}{27}\sin^2\left(\frac{3\theta}{2}
	\right)t^{-4},
	\end{align}
	or, equivalently, in terms of $a$,
		\begin{align}
	\label{kretschmann_singularity_a}
	K\propto a^{-12}.
	\end{align}
	As we can observe from \eqref{kretschmann_singularity_cosmological}, in terms of the cosmological time, the form of the Kretschmann scalar in this limit does
	not depend on the matter content. In fact, its scaling follows
	the same power law as in the large-volume limit \eqref{kretschmann_big_volumes_tau}.
	But, since the singularity is located at $t=0$, in this case the scalar will diverge.

	Note that in the above expressions \eqref{kretschmann_singularity}--\eqref{kretschmann_singularity_a}, we have
	only displayed the leading term, which, if $\sin(3\theta/2)=0$, will be vanishing. However,
	under the presence of matter, there are always subdominant diverging terms,
	and the Kretschmann scalar generically diverges.
	The vacuum scenario is the only case for which there are certain trajectories
	that have a constant and exactly vanishing Kretschmann scalar,
	which thus does not diverge towards the singularity. It is easy to see
	this feature from its exact form \eqref{Kretschmann_vacuum}: if
	either $p_+=-P$ (and thus $\theta=0$) or $p_+=P/2$ (and thus $\theta=2\pi/3$ or $\theta=4\pi/3$), then $K=0$.
	These special trajectories correspond to the following three sets of Kasner exponents \eqref{def_Kasner_exp}: $(p_1=p_2=0,p_3=1)$, $(p_1=p_2=0,p_2=1)$, and $(p_2=p_3=0,p_1=1)$, which define the only three possibilities with all the
	Kasner exponents being nonnegative, and it implies that
	none of the scale factors \eqref{evol_Kasner} expand towards the singularity.

Finally, concerning the Weyl square $W^2$ in the limit
towards the singularity, in \eqref{Weyl_singularity} it is explicit that
the term related to the Ricci scalar and the Ricci square scales as $a^{-6(1+\omega)}$. If $\omega\in[-1,1)$, this term diverges slower than the Kretschmann scalar \eqref{kretschmann_singularity_a} and, hence, the latter will dominate.
For stiff matter $\omega=1$, both terms show the same divergence,
and thus both should be taken into account to compute $W^2$.
In vacuum, the Ricci tensor is exactly vanishing, one has that
$W^2=K$, and, in particular, $W^2$ will also be vanishing along the
special trajectories
commented above.

\section{Conclusions}\label{sec:conclusions}

We have obtained exact analytic solutions
to describe the evolution of the Bianchi I geometry coupled to several
perfect fluid species.
The dynamical variables that describe the evolution of the model
are the energy density of each species $\rho_s$, the two shape
parameters $\beta_{\pm}$, which encode the anisotropy,
and the average scale factor $a$, which is defined as the geometric
average of the scale factors in the different spatial directions.

In the chosen gauge,
the evolution of the matter energy density $\rho_s$ and the shape parameters $\beta_{\pm}$
can be analytically obtained in general. Furthermore, the average scale factor $a$ follows the Friedmann
equation \eqref{eq_friedmann}, where the anisotropies are encoded in a constant
term $P^2=(p_+^2+p_-^2)$, which
can be understood as the (square of the) momentum of a massless scalar field.
The solution to this equation can be reduced to a quadrature,
and we have described the qualitative evolution of the system
by understanding the Friedmann equation as an energy equation
for a particle moving on a potential [see Eq. \eqref{eq_conservation_V}]. In addition, we have obtained
the explicit analytic solution
for a number of cases. In particular,
we have derived the Kasner solution for vacuum \eqref{sol_a_Kasner_vacuum}, as well
as the solution for a single fluid species \eqref{sol_a_single_matter_pos}. Although this
solution has been studied in the literature, we have not found it
anywhere given in such explicit and compact form as here.
We have also considered the case with a negative
energy density \eqref{sol_a_single_matter_neg}, which includes the model with a negative
cosmological constant, and generically produces a recollapse
of the universe, leading to a cosmology with an initial and a final
singularity. Concerning the case with two fluid species,
the solution is obviously more complicated, but,
if the barotropic indices obey certain relation, the integral
can be performed and the solution can be explicitly written.
Such relation includes the particular cases of a Bianchi I universe either
with cosmological constant and dust, or with
radiation and cosmic strings. For completeness, we have
obtained the explicit solution \eqref{sol_bianchi_I_k=2_pos}--\eqref{sol_bianchi_I_k=2_neg_special} for all the different signs
of the energy densities. Exotic species with a negative
sign of the density can produce turning points of the scale factor
and bound (either from above or below) its possible values.
Among the different possibilities, we find an interesting case,
with an exotic and nonexotic component, which leads to a singularity-free cosmology: the universe collapses from infinite volume
until a certain minimum value of the scale factor, where a bounce
happens, and then it expands forever. The exotic component
is dominant around the bounce, while, for large volumes,
it decays and becomes negligible as compared to the nonexotic one.
We consider that this, and similar models, could be worth studying
as an effective description of a nonsingular cosmology.

Finally, we have analyzed the asymptotic behavior of the Bianchi I
universe for both large volumes and towards the singularity. In such limits,  we have studied in detail the scaling of the different curvature invariants,
and explicitly derive their dependence on the different parameters of the model. For large volumes, the anisotropies decay and the universe
isotropizes tending to a FLRW geometry. Towards the singularity,
we have obtained the well-known scaling of the different
scalars that show the blowup of the curvature.

\section*{Acknowledgments}
This work is supported by
the Basque Government Grant No.~\mbox{IT1628-22}, and
by the Grant No.~PID2021-123226NB-I00 (funded by MCIN/AEI/10.13039/501100011033 and by “ERDF A way of making Europe”).
S.F.U. is funded by an FPU fellowship of the Spanish Ministry of Universities.

\appendix

\section{Plots for two species with $1-\omega_-=2(1-\omega_+)$}

Here we present certain plots for the case with two species whose barotropic indices satisfy $1-\omega_-=2(1-\omega_+)$.
The figures show the qualitative form of the potential $V(a)$,
defined in \eqref{def_V_a},
as well
as the evolution \eqref{sol_bianchi_I_k=2_pos}--\eqref{sol_bianchi_I_k=2_neg_special} of the average scale factor $a=a(\tau)$,
for different values of the densities $\rho_{01}$ and $\rho_{02}$.

\begin{figure}[h]
		\centering
	\begin{minipage}{0.461\textwidth}
		\centering
	\includegraphics[width=0.95\linewidth]{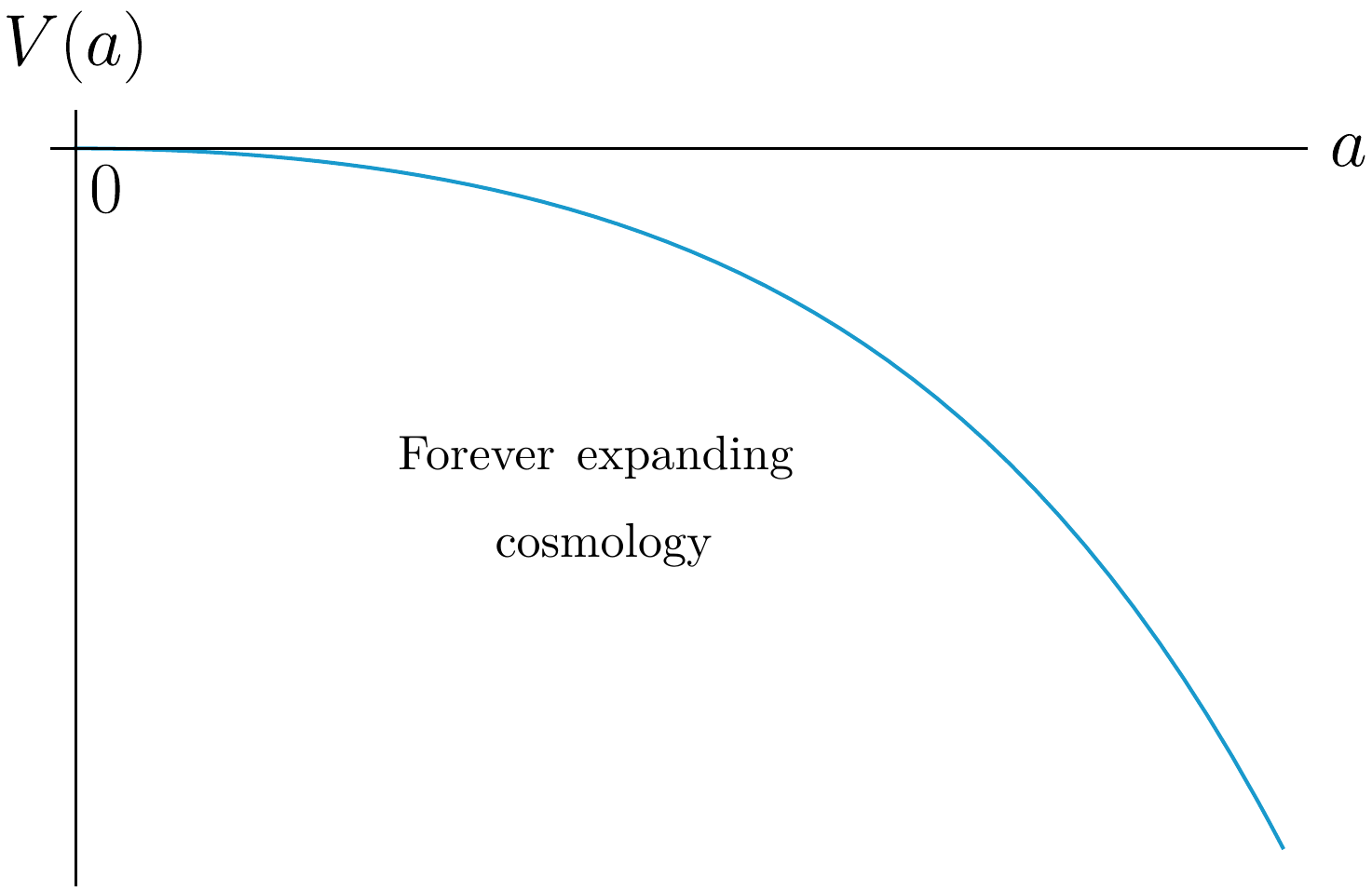}
	\caption{Shape of $V(a)$ for two species whose barotropic indices satisfy $1-\omega_2=2(1-\omega_1)$, with positive densities $\rho_{01},\rho_{02}>0$.
	In this scenario, $V(a)$ has no positive roots, and thus it corresponds to a forever expanding cosmology
	with $a\in(0,+\infty)$.
	}
	\label{fig:v_2_pos}
	\end{minipage}
	\hfill
		\begin{minipage}{0.461\textwidth}
		\centering
		\includegraphics[width=0.95\linewidth]{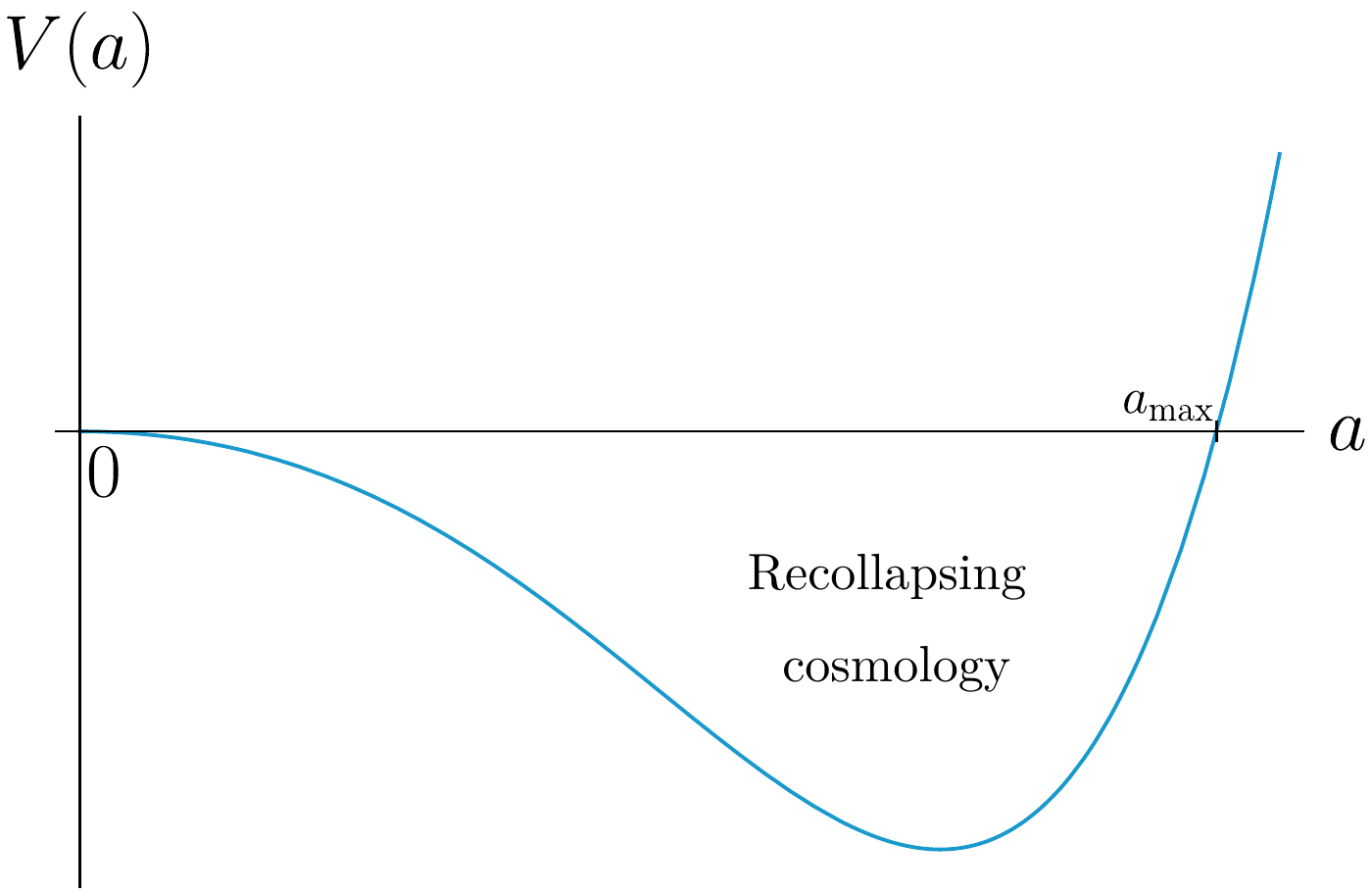}
		\caption{Shape of $V(a)$ for two species whose barotropic indices satisfy $1-\omega_2=2(1-\omega_1)$, with negative density $\rho_{02}<0$.
When $a\to 0$, the first species dominates, while for $a\to +\infty$, the second one dominates,
making $V(a)$ to take positive values.
Consequently,
$V(a)$ has a single and positive root, $a_{\max}$. This 
		scenario represents thus a recollapsing cosmology with $a\in (0,a_{\rm max}]$.
		}
		\label{fig:v_2_rho_-_neg}
	\end{minipage}
	\vfill
\vspace{1cm}
	\begin{minipage}{0.461\textwidth}
		\centering
		\includegraphics[width=0.95\linewidth]{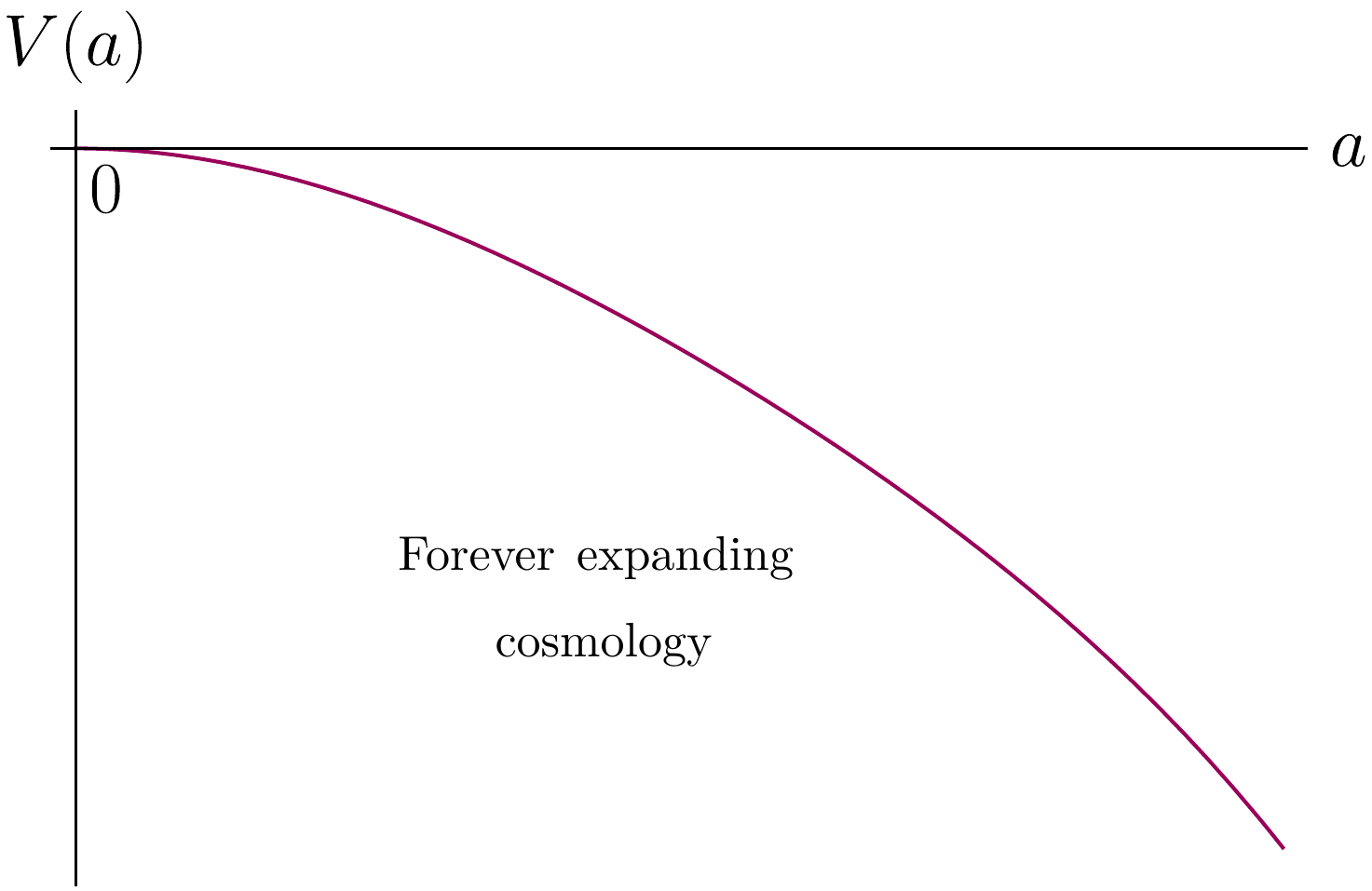}
		\caption{Shape of $V(a)$ for two species whose barotropic indices satisfy $1-\omega_2=2(1-\omega_1)$, with negative density $\rho_{01}<0$ and positive density $\rho_{02}>\rho_{\rm threshold}$. In this case, $V(a)$ has no real roots and thus it represents a forever expanding cosmology with $a\in(0,+\infty)$.}
		\label{fig:v_2_rho_-_pos_u_thres}
	\end{minipage}
	\hfill
	\begin{minipage}{0.461\textwidth}
	\centering
	\includegraphics[width=0.95\linewidth]{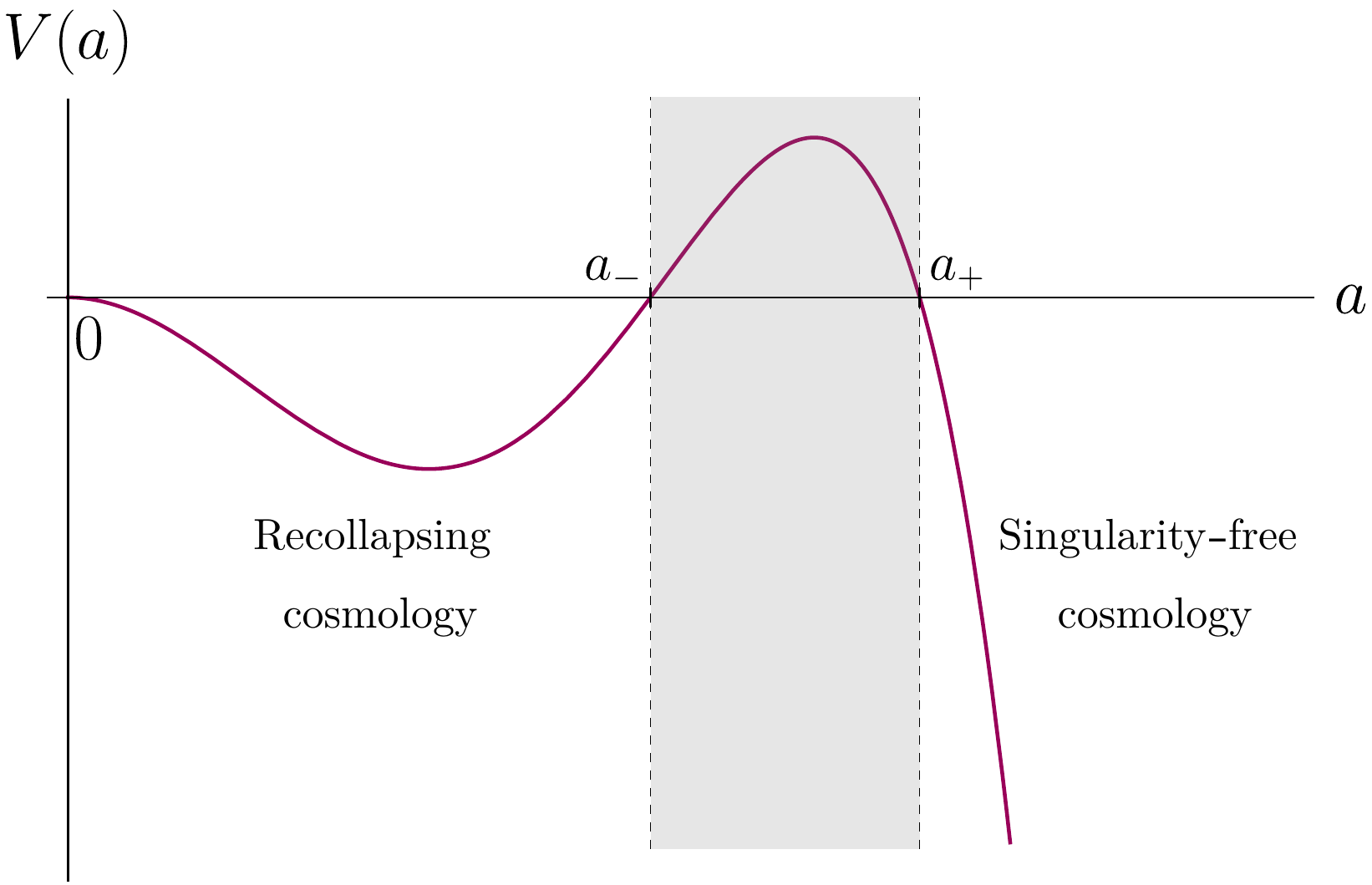}
	\caption{Shape of $V(a)$ for two species whose barotropic indices satisfy $1-\omega_2=2(1-\omega_1)$, with negative density $\rho_{01}<0$ and positive density $0\leq\rho_{02}<\rho_{\rm threshold}$. In this case, $V(a)$ has two positive roots, $a_{\pm}$, which define two regions with different cosmological evolutions:
	a recollapsing cosmology with $a\in(0,a_-]$, and a singularity-free cosmology that expands towards infinitely large volumes with $a\in[a_+,+\infty)$.}
	\label{fig:v_2_rho_-_pos_d_thres}
\end{minipage}
\end{figure}

\begin{figure}[h!]
	\centering
	\begin{minipage}{0.465\textwidth}
		\centering
		\includegraphics[width=\linewidth]{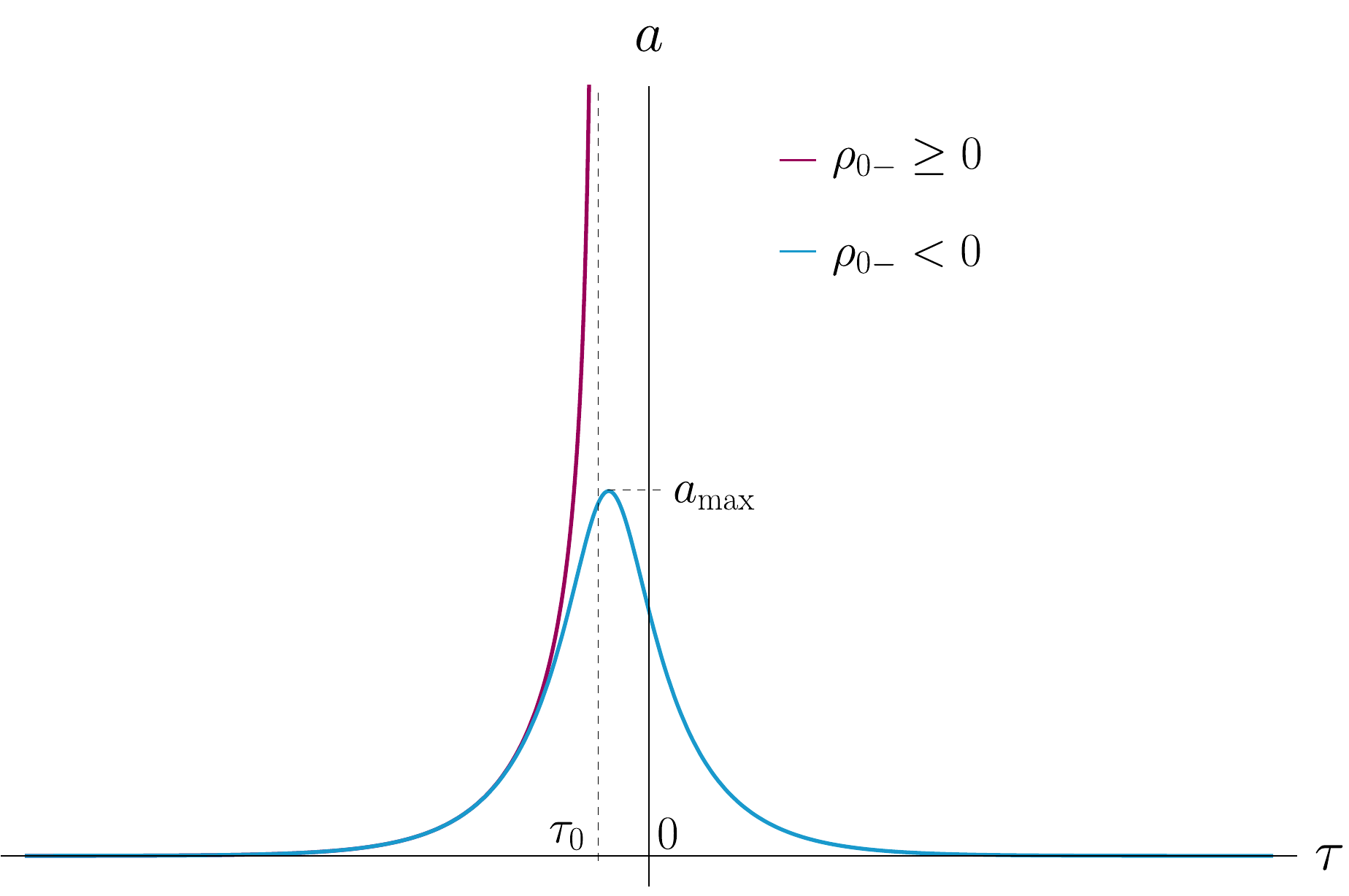}
		\caption{Evolution \eqref{sol_bianchi_I_k=2_pos} for $a=a(\tau)$, corresponding to the scenario where there are
		two species with barotropic indices satisfying $1-\omega_-=2(1-\omega_+)$ and positive density $\rho_{0+}>0$. Each curve corresponds to a sign of $\rho_{0-}$.
		Depending on this sign, the cosmological evolution presents an infinite expansion
		(for $\rho_{0-}\geq 0$) or a recollapse (for $\rho_{0-}<0$). As can be seen, in the limit towards the singularity ($\tau\to -\infty$),
		both evolutions match, as it corresponds to the region where $\rho_{0+}$ dominates, and thus the value of $\rho_{0-}$ is irrelevant.
		}
		\label{fig:evol_a_rho01_pos}
	\end{minipage}
	\hfill
	\begin{minipage}{0.465\textwidth}
		\centering
		\includegraphics[width=\linewidth]{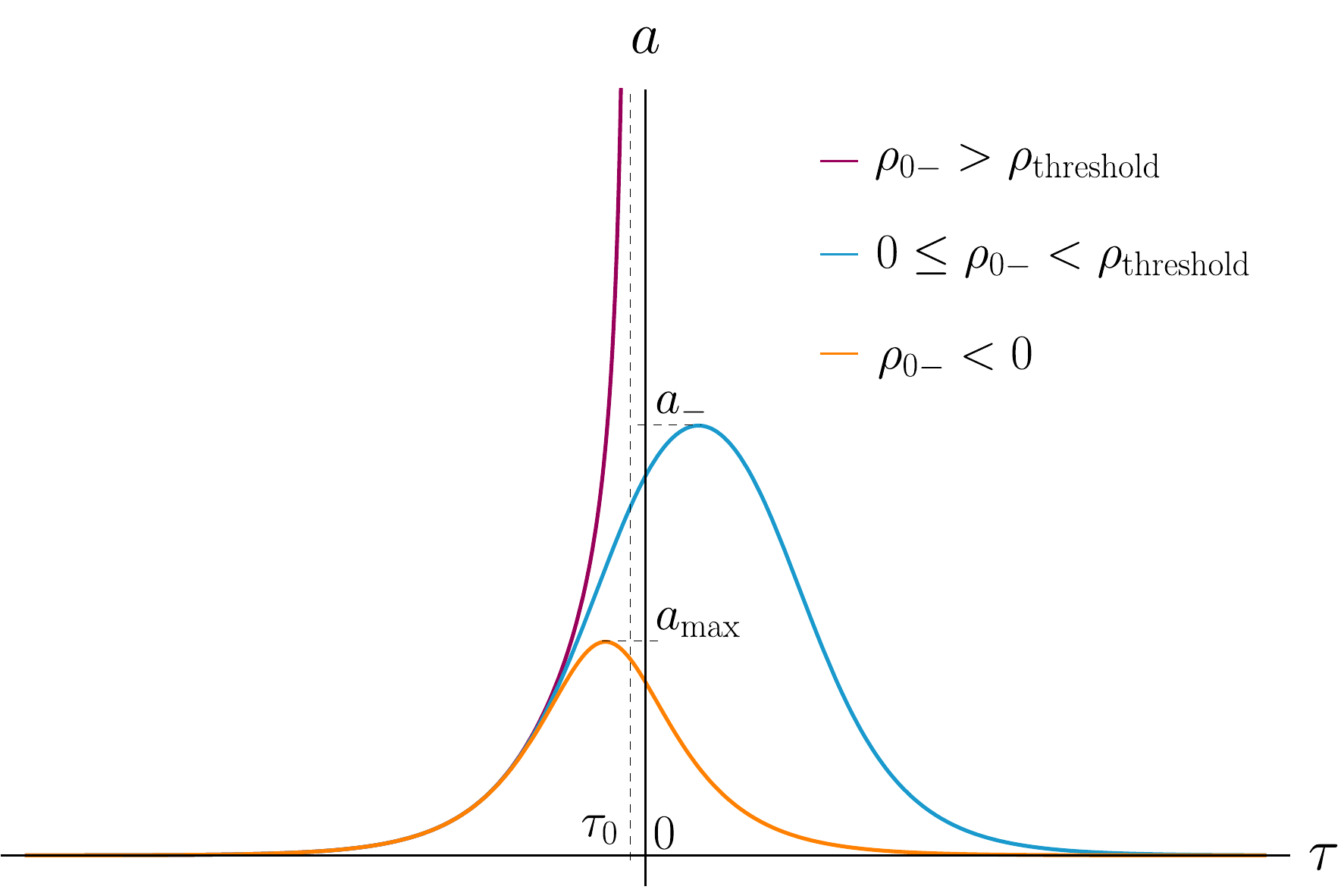}
		\caption{Evolution \eqref{sol_bianchi_I_k=2_neg} for $a=a(\tau)$, corresponding to the scenario where there are two species satisfying $1-\omega_-=2(1-\omega_+)$
		and with negative density $\rho_{0+}<0$. Each curve corresponds to a different value of $\rho_{0-}$.
		Depending on the latter, there are different cosmological evolutions, presenting either an infinite expansion (for $\rho_{0-}>\rho_{\rm threshold}$)
		or a recollapse (for $\rho_{0-}<\rho_{\rm threshold}$).
		As can be  observed, in the limit towards the singularity $\tau\to -\infty$, all three evolutions match,
		as it corresponds to the region where $\rho_{0+}$ dominates, and thus the value of $\rho_{0-}$ is irrelevant.
		}
		\label{fig:evol_a_rho01_neg_gen}
	\end{minipage}
	\vfill
	\vspace{1cm}
	\begin{minipage}{0.465\textwidth}
		\centering
		\includegraphics[width=\linewidth]{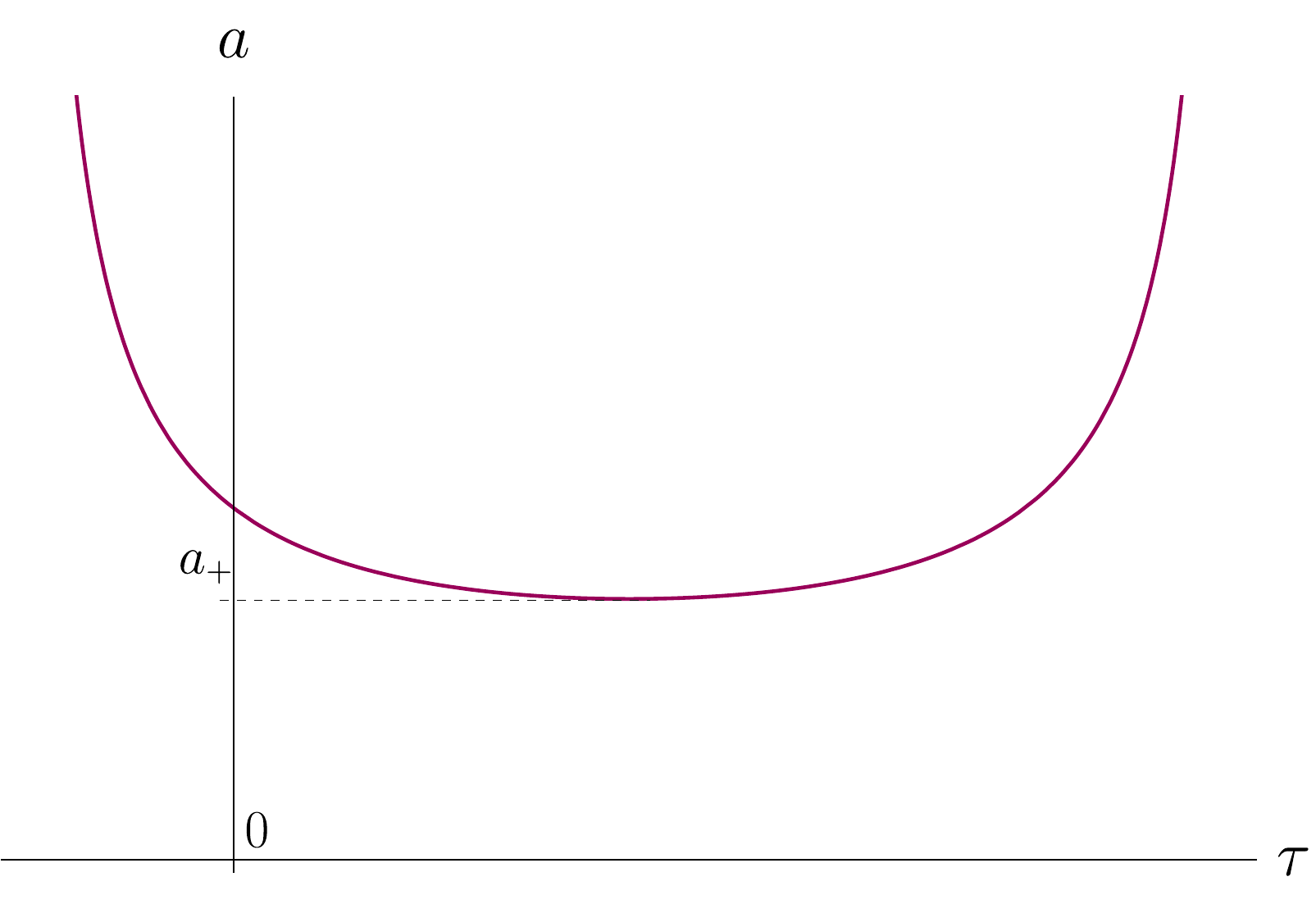}
		\caption{Evolution \eqref{sol_bianchi_I_k=2_neg_special} for $a=a(\tau)$, corresponding to the scenario where there are two species
		whose barotropic indices satisfy $1-\omega_-=2(1-\omega_+)$, with negative density $\rho_{0+}<0$ and for $0\leq\rho_{0-}<\rho_{\rm threshold}$.
		As can be seen, this evolution corresponds to a singularity-free cosmology, with a bounce at $a=a_+$.}
		\label{fig:evol_a_rho01_neg_special}
	\end{minipage}
	\hfill
\begin{minipage}{0.465\textwidth}
	\centering
	\includegraphics[width=\linewidth]{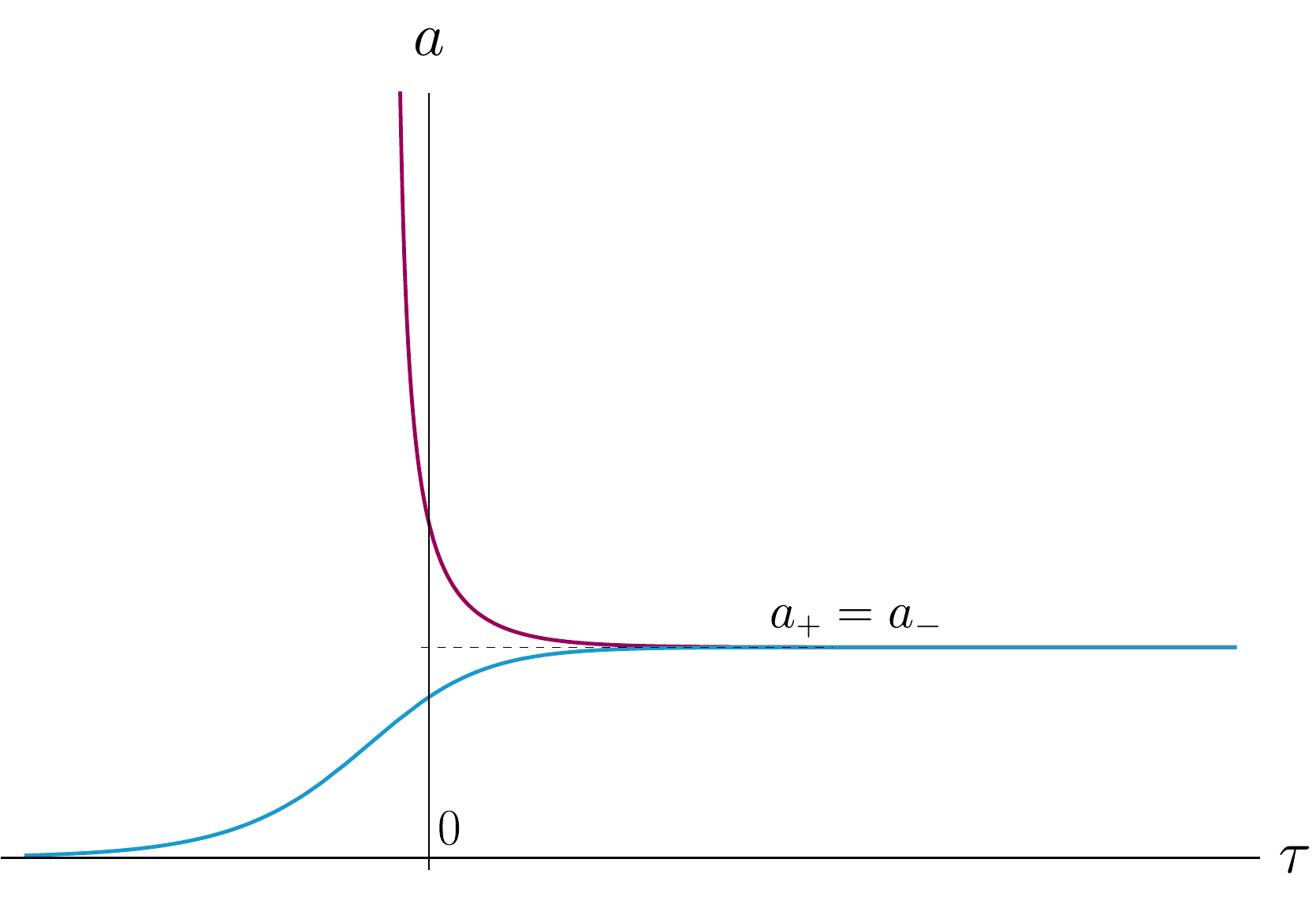}
		\caption{Evolution of $a=a(\tau)$, corresponding to the degenerate case with two species satisfying $1-\omega_-=2(1-\omega_+)$,
	negative density $\rho_{0+}>0$, and $\rho_{0-}=\rho_{\rm threshold}$.
	The singularity-free cosmological evolution, given by \eqref{sol_bianchi_I_k=2_neg} and shown in purple,
	begins with an infinite value of $a$ and asymptotically tends to the minimum $a=a_-=a_+$.
	The other independent evolution, given by \eqref{sol_bianchi_I_k=2_neg_special} and shown in blue,
	corresponds to an ever-expanding universe asymptotically tending to the maximum $a=a_-=a_+$.}
	\label{fig:evol_a_rho01_deg}
\end{minipage}
\end{figure}

\vspace{5cm}

\bibliographystyle{bib-style}
\bibliography{references}

\end{document}